\RequirePackage{rotating}
\documentclass[fleqn,usenatbib]{mnras}

\usepackage{newtxtext,newtxmath}

\usepackage[T1]{fontenc}

\DeclareRobustCommand{\VAN}[3]{#2}
\let\VANthebibliography\thebibliography
\def\thebibliography{\DeclareRobustCommand{\VAN}[3]{##3}\VANthebibliography}

\usepackage{graphicx}	
\usepackage{amsmath}	

\defcitealias{Huitson2017}{H17}
\defcitealias{Panwar2022}{P22}

\title[Gemini/GMOS transmission spectrum of WASP-19b]{A new method to correct for host star variability in multi-epoch observations of exoplanet transmission spectra} 

\author[Panwar et al.]{
Vatsal Panwar,$^{1}$\thanks{E-mail: v.panwar@uva.nl}
Jean-Michel D\'esert,$^{1}$\thanks{E-mail: desert@uva.nl}
Kamen O. Todorov,$^{1}$
Jacob L. Bean,$^{2}$
Kevin B. Stevenson,$^{3}$
\newauthor
C. M. Huitson,$^{4}$
Jonathan J. Fortney,$^{5}$
and Marcel Bergmann$^{6}$
\\
$^{1}$Anton Pannekoek Institute for Astronomy,
University of Amsterdam, P.O. Box 94249, 1090GE Amsterdam, Noord Holland, NL\\
$^{2}$Department of Astronomy and Astrophysics, University of Chicago, Chicago, IL 60637, USA\\
$^{3}$JHU Applied Physics Laboratory, 11100 Johns Hopkins Rd, Laurel, MD 20723, USA \\
$^{4}$CASA, University of Colorado, 389 UCB, Boulder, CO, 80309-0389, USA \\
$^{5}$Department of Astronomy and Astrophysics, University of California, Santa Cruz, CA 95064, USA \\
$^{6}$NOAO, Gemini Observatory, present address Palo Alto, CA, USA
}

\date{Accepted 2022 July 4. Received 2022 June 30; in original form 2022 January 23}

\pubyear{2021}

\begin{document}
\label{firstpage}
\pagerange{\pageref{firstpage}--\pageref{lastpage}}
\maketitle

\begin{abstract}

Transmission spectra of exoplanets orbiting active stars suffer from wavelength-dependent effects due to stellar photospheric heterogeneity. WASP-19b, an ultra-hot Jupiter (T$_{eq}$ $\sim$ 2100 K), is one such strongly irradiated gas-giant orbiting an active solar-type star. We present optical (520-900 nm) transmission spectra of WASP-19b obtained across eight epochs using the Gemini Multi-Object Spectrograph (GMOS) on the Gemini-South telescope. We apply our recently developed Gaussian Processes regression based method to model the transit light curve systematics and extract the transmission spectrum at each epoch. We find that WASP-19b's transmission spectrum is affected by stellar variability at individual epochs. We report an observed anticorrelation between the relative slopes and offsets of the spectra across all epochs. This anticorrelation is consistent with the predictions from the forward transmission models, which account for the effect of unocculted stellar spots and faculae measured previously for WASP-19. We introduce a new method to correct for this stellar variability effect at each epoch by using the observed correlation between the transmission spectral slopes and offsets. We compare our stellar variability corrected GMOS transmission spectrum with previous contradicting MOS measurements for WASP-19b and attempt to reconcile them. We also measure the amplitude and timescale of broadband stellar variability of WASP-19 from TESS photometry, which we find to be consistent with the effect observed in GMOS spectroscopy and ground-based broadband photometric long-term monitoring. Our results ultimately caution against combining multi-epoch optical transmission spectra of exoplanets orbiting active stars before correcting each epoch for stellar variability.


\end{abstract}



\section{Introduction}
\label{sec_intro}


Ground-based observations using low resolution multi-object spectroscopy (hereafter referred to as MOS) on large telescopes (\citealt{Bean2010}, \citeyear{Bean2011}) have yielded precise optical and near-infrared transmission spectra which have helped to constrain the atmospheric properties of exoplanets ranging from transiting hot gas giants (e.g. \citealt{Nikolov2018}) to smaller and cooler rocky exoplanets (e.g. \citealt{Diamond-Lowe2018}). Conventionally, the ground-based MOS technique has been restricted to exoplanets transiting host stars with comparison stars of similar brightness and spectral type in the instrument's field of view, which can be used for differential spectrophotometry. However, recent development in the techniques of modelling telluric and instrumental systematics in this context (\citealt{Panwar2022}, hereafter referred to as \citetalias{Panwar2022}) have also extended the application of the MOS technique to exoplanets orbiting host stars, including bright stars targeted by TESS planet detection campaigns, with no suitable nearby comparison stars.  

A long-standing issue in transit spectroscopy of exoplanets has been the contamination of the planetary spectrum due to stellar variability stemming from stellar photospheric heterogeneity. The amplitude of such contamination can be comparable to the desired precision of the transmission spectra (e.g. \cite{Rackham2017GJ1214}). The level of contamination is particularly significant for active solar type stars observed in the optical wavelength range typically probed by ground-based MOS observations using instruments like VLT/FORS2 or Gemini/GMOS. This wavelength dependent effect (\citealt{Pont2008}, \citealt{McCullough2014}) stems from unocculted or occulted magnetically active regions like spots and faculae in the stellar photosphere and has recently come to be commonly referred to as the transit light source effect (\citealt{Rackham2018}). Several observations and in-depth modelling (\citeauthor{Rackham2017}, \citeyear{Rackham2018}, \citeyear{Rackham2019}) have revealed this wavelength-dependent effect as imprinted on the transmission spectra of exoplanets orbiting active stars (e.g. \citealt{Espinoza2019}, \citealt{Kirk2021}, \citealt{Sedaghati2021}, \citealt{Nikolov2021}).  

The transit light source effect has been observed and modelled in the MOS observations of many transiting exoplanets in recent years. The framework of (\citeauthor{Rackham2017}, \citeyear{Rackham2018}, \citeyear{Rackham2019}) was first implemented in a Bayesian atmospheric retrieval code \texttt{AURA} by \cite{Pinhas2018} and was recently also used by \cite{Nikolov2021} to model the transit light source effect in the transmission spectrum of WASP-110b. Other Bayesian retrieval codes like \texttt{POSEIDON} introduced by \cite{MacDonald2017} (e.g. applied to WASP-103b observations in \citealt{Kirk2021}) and \texttt{platon} (e.g. applied to the VLT/FORS2 observations of WASP-19b in \cite{Sedaghati2021}) also fit for the stellar photospheric heterogeneity parameters together with the planetary atmosphere.     

WASP-19b (\citealt{Hebb2010}) is one such example of a transiting gas giant exoplanet orbiting an active G dwarf with significant stellar variability. Active FGK dwarfs have been known to produce prominent features in the transmission spectra (\citealt{Rackham2019}) due to stellar activity. Hence, it is pertinent to account for the effect of stellar activity when studying the atmosphere of WASP-19b. WASP-19b also falls in the class of ultra-hot Jupiters (T$_{eq}$ $\gtrsim$ 2000 K; e.g., \citealt{Arcangeli2018}, \citealt{Lothringer2018},\citealt{Kitzmann2018}), which have recently been the subject of atmospheric characterization in the optical through low resolution MOS (e.g. \citealt{Stevenson2014}, \citealt{Wilson2021}) and high resolution spectroscopy (e.g. \citealt{Hoeijmakers2019}, \citealt{Pino2020}, \citealt{Ehrenreich2020}).  

Studies presenting discrepant optical transmission spectrum of WASP-19b have been published recently using VLT/FORS2 (\citealt{Sedaghati2017}), VLT/ESPRESSO (\citealt{Sedaghati2021}), and Magellan/IMACS by \cite{Espinoza2019}. This motivated us to follow up the system using Gemini/GMOS. 

The paper is distributed as follows. We first review the state of the art in atmospheric studies of WASP-19b in Section \ref{sec:intro_wasp19b}. In Section \ref{sec_obs} we describe our observations of WASP-19b from Gemini/GMOS. In Section \ref{sec_data_reduc} we describe our data reduction of these observations and in Section \ref{sec_analysis} we discuss the analysis of transit light curves to obtain the transmission spectrum. In Section \ref{sec:discussion} we discuss the interpretation of the transmission spectrum, especially in the context of the host star's activity. We also compare the results from GMOS observations with forward transmission spectrum models accounting for the effect of stellar variability. We introduce a new empirical approach to correct for the effect of stellar variability in the transmission spectrum at individual epochs before constructing the final combined transmission spectrum. We discuss the implications for the atmosphere of WASP-19b from the combined GMOS and HST/WFC3 transmission spectrum  in Section \ref{sec:atmosphere_interpretation}. We further put in context the effect of stellar variability observed in the broadband transit depths measured from TESS photometry of 58 transits of WASP-19b observed over two sectors. We describe our analysis of TESS and ground based photometric follow-up from Las Cumbres Observatory Global to monitor the stellar variability of WASP-19 in the Appendix \ref{app:photometric_monitoring}. Specifically, we use the long-term photometry of the system from TESS covering several transits to understand the effect of stellar variability on the broadband optical transit depth, and compare it with the relative variations seen between the GMOS transmission spectra at multiple epochs. In Section \ref{conclusions} we present our conclusions.

\section{The case of WASP-19b}
\label{sec:intro_wasp19b}
WASP-19b, one of the shortest period Jupiter mass gas giant exoplanets known (orbiting a G8V star in just 18.9 hours), is situated in the "sub-Jupiter" desert in the mass vs orbital period distribution of the population of hot Jupiters which shows a pileup around orbital period of 3 -- 4 days (\citealt{SzaboKiss2011}, \citealt{Hellier2011}). It is also an ideal candidate for atmospheric characterization on multiple accounts. With the high level of stellar irradiation and resultant equilibrium temperature of 2100 K, and low surface gravity (log$_{10}g [m/s^{2}] = $ 2.15), WASP-19b is expected to have TiO and VO at gas phase equilibrium in the upper atmosphere that, if present in a cloud free atmosphere, will absorb the incident optical stellar flux and could cause thermal inversion (\citealt{Hubeny2003}, \citealt{Fortney2008}). The host star WASP-19 is also known to be active, with the optical stellar flux varying peak to trough 2 to 3 \% at a period of $\sim$ 10.5 days (\citealt{Hebb2010}, \citealt{Huitson2013}, \citealt{Espinoza2019}). The chromospheric Ca II H \& K line emission ratio of WASP-19 quantified by log(R$'_{HK}$) = -- 4.5 $\pm$ 0.03 (\citealt{Anderson2013}, \citealt{Knutson2010}) quantifies the high level of chromospheric activity of the star. Table \ref{tab:stellar_params} shows the properties of the host star WASP-19 from the literature.

With a dayside temperature of 2240 $\pm$ 40 K (inferred from TESS and previous secondary eclipse depth measurements, \citealt{Wong2019}), WASP-19b is on the cusp of transition of hot to ultra-hot Jupiters (\citealt{Parmentier2018}, \citealt{Baxter2020}), at which point atmospheric opacities, molecular dissociation, H$-$ opacity, latent heat and thermal inversion begin to become relevant (\citealt{Arcangeli2018}, \citealt{Lothringer2018}, \citealt{Kitzmann2018}). Retrieval analysis of emission spectra including secondary eclipse depth measurements from Spitzer and TESS secondary eclipse observations (\citeauthor{Wong2016} \citeyear{Wong2016}, \citeyear{Wong2019}) indicate an atmosphere with no dayside thermal inversion and moderately efficient day-night circulation. However, in contrast to these findings, \cite{Rajpurohit2020} interpret the excess eclipse depth in the Spitzer 4.5 $\mu$m band as due to CO in emission and thus as an evidence of thermal inversion in the atmosphere of WASP-19b.

Using transmission spectroscopy of WASP-19b, \cite{Huitson2013} have detected absorption features due to water in the 1.1-1.7 $\mu$m range HST/WFC3 G141 observations, which is consistent with a solar abundance atmosphere with no or only low level of clouds. There is evidence that high levels of UV flux from active stars could be responsible for the dissociation of molecular absorbers like TiO (\citealt{Knutson2010}). \cite{Huitson2013} hypothesize this to be one of the possible reasons behind non-detection of TiO in their HST/STIS optical transmission spectrum. The presence or absence of TiO in the atmosphere can affect the overall energy budget of WASP-19b, drives thermal inversion in the atmosphere, and ultimately affects the inferences about the atmospheric metallicity and C/O which hold potential clues to the formation and evolution history of gas giants (\citealt{Madhusudhan2012}, \citealt{Mordasini2016}, \citealt{Eistrup2018}). 


The picture in the optical wavelength range of the transmission spectrum of WASP-19 is mired with a discrepancy due to two different studies reporting contrasting results. \cite{Sedaghati2017} from their observations obtained using VLT/FORS2 first reported the detection of TiO features in the optical transmission spectrum with a strong scattering slope due to hazes towards the blue end and a water feature towards the red end at high significance. However, \cite{Espinoza2019} detect a featureless optical transmission spectrum from their observations using Magellan/IMACS, with no significant TiO features and no slope due to hazes. This is consistent with the picture apparent from low resolution optical transmission spectrum from HST/STIS reported by \cite{Huitson2013}. \cite{Sedaghati2021} use high resolution spectroscopic observations from VLT/ESPRESSO to search for signatures of atomic and molecular species in the optical via cross-correlation analysis and report a tentative indication of TiO at $\sim$ 3 $\sigma$ confidence. Through chromatic, Rossiter-McLaughlin effect analysis \cite{Sedaghati2021} also report a strong scattering slope towards the blue wavelengths, consistent with the findings of \cite{Sedaghati2017} at low-resolution and in contrast with the flat spectrum presented by \cite{Espinoza2019}. 

Activity and variability of the host star WASP-19 contaminates the transmission spectrum of the planet via the transit light source effect (\citealt{Rackham2017}). \cite{Espinoza2019} observed occultations of stellar spots and plages and used them to put constraints on the spot size and spot temperature contrast with respect to the stellar photosphere. Interestingly, the transmission spectrum from one of the six epochs analysed by \cite{Espinoza2019} shows a significantly steeper slope compared to those from other epochs due to stellar activity. \cite{Espinoza2019} perform retrievals accounting for stellar activity on the transmission spectra from all epochs independently. They find that the epoch showing a steep slope can be best explained by strong stellar contributions from stellar activity. However, all the other five epochs show no statistically significant contribution from stellar activity contamination and are most consistent with a flat line. \cite{Espinoza2019} eventually reject the spectrum with steep slope when they construct the combined transmission spectrum from the mean subtracted transmission spectra of the other five epochs. They also do not apply any additional slope corrections to the individual spectra before combining them. 

\cite{Sedaghati2021} in their reanalysis of the VLT/FORS2 observations of \cite{Sedaghati2017} analyse the effect of stellar surface heterogeneity on WASP-19b's transmission spectrum through a \texttt{POSEIDON} (\citealt{MacDonald2017}) retrieval analysis of the transmission spectra from the three epochs. Each VLT/FORS2 epoch was observed in a different wavelength range, going from blue to red optical. \cite{Sedaghati2021} from their retrieval analysis find that the VLT/FORS2 spectrum is best explained by an atmosphere with 100$\times$ sub-solar TiO. They also find that after accounting for stellar activity, the significance of TiO detection in the VLT/FORS2 spectrum goes from 7.7$\sigma$ to 4.7$\sigma$. The stellar spot contrast and covering fractions retrieved by \cite{Sedaghati2021} from their VLT/FORS2 spectrum are consistent with those measured by \cite{Espinoza2019} from their Magellan/IMACS spectrum. Additionally, \cite{Sedaghati2021} also perform a retrieval on the Magellan/IMACS combined transmission spectrum from \cite{Espinoza2019} and find a marginal preference for the model with TiO ($\Delta$lnZ = 0.5) compared to a flat line or models with only contributions from stellar activity. 

In summary, both FORS2 and IMACS spectra have confirmed the significant effect of stellar activity in the transmission spectrum of WASP-19b. Both spectra have different morphologies and an agreement between them still remains at a marginal threshold as indicated by the retrieval of the IMACS spectrum by \cite{Sedaghati2021}. This tension in the observations of WASP-19b's atmosphere, including the presence or absence of TiO, motivated us to further investigate its optical transmission spectrum, which we present in this paper. In this paper, we present a study of WASP-19b's transmission spectrum from 8 epochs observed using Gemini/GMOS in the wavelength range of 520 to 900 nm. We present a new approach to analyse and correct the effect of stellar variability at each epoch by looking at its two broad manifestations: the slope and the offset of the transmission spectrum. The new data analysis method introduced in \citetalias{Panwar2022} mitigates potential systematics due to non-linear differences between the target and comparison star light curves and enables accurate measurement of the slopes and offsets of the transmission spectrum at each epoch.

\begin{table}
\caption[1]{Stellar parameters of WASP-19.}
\label{tab:stellar_params}
\centering

\begin{tabular}{lll}
\hline
\hline
M$_\star$ (M$_{\odot}$) & $0.904\pm0.04$ & \cite{Tregolan-Reed2013} \\
R$_\star$ (R$_{\odot}$) & $1.001\pm0.035$ & \cite{Gaia2018}) \\
P$_{\rm rotation}$ (days) & $10.50\pm0.2$ & \cite{Bonomo2017} \\
$\rm V_\star$ (mag)   &  $12.31\pm0.04$ & \cite{Zacharias2013} \\
T$_{eff,\star}$ (K) &    $5460^{+90}_{-90}$ & \cite{Doyle2013} \\  
SpT$_\star$ & G8V & \cite{Hebb2010} \\
L$_\star$ (log$_{10}$L$_{\odot}$) & $-0.09\pm0.005$ & \cite{Gaia2018} \\
$\rm \log(g_\star$) & $4.45\pm0.05$ & \cite{Torres2012} \\\relax
[Fe/H]$_\star$ & $0.15\pm0.07$ & \cite{Torres2012} \\
Distance (pc) & $270.41\pm1.46$ & \cite{Gaia2018} \\
log(R$'_{HK}$) & -- 4.5 $\pm$ 0.03 & \cite{Anderson2013} \\
\hline
\end{tabular}
\label{tab:stellar_prop}
\end{table}


\section{Multi-epoch transit observations of WASP-19b}
\label{sec_obs}


\subsection{Gemini/GMOS transit observations of WASP-19b}
\label{gmos_obs}
We observed eight transits of WASP-19b (Table \ref{obsstats}) in the red optical using GMOS on the Gemini South telescope located at Cerro Pachon, Chile. Since the host star WASP-19 is known to be active (e.g. spot crossing events seen in the observations by \citealt{Espinoza2019}), we spread the observations over a period of 2 years. All 8 transits were observed as part of a survey program of hot Jupiter atmospheres from Gemini/GMOS (Proposal ID: 2012B-0398; PI: J-.M D\'{e}sert) and described in more detail in \cite{Huitson2017} (referred to as \citetalias{Huitson2017} hereafter). The observations were performed using the same set-up as described in \citetalias{Huitson2017}, which is similar to that of previous exoplanet atmospheric observations using GMOS (e.g. \citeauthor{Bean2010} \citeyear{Bean2010}, \citeyear{Bean2011}; \citealt{Gibson2013}). For each observation, we used the multi-object spectroscopy mode of GMOS-South to obtain time series spectrophotometry of WASP-19 and 2 nearby comparison stars (described in more detail below) simultaneously. All the 8 transits were observed in the red optical using R150 + G5306 grating combination, covering a wavelength range of 525-900~nm with an ideal resolving power $R \sim 600$. The ideal resolving powers assume a slit width of 0.5~arcsec. We used masks with 10~arcsec wide slits on each star, and obtained a seeing-limited spectral resolution. Given the range of seeing measured during our observations ( Table~\ref{obsstats}) our resolution is approximately $2-3\times$ lower than the ideal value depending on observation. 

For all the observations, we used the grating in first order. The requested central wavelength was 620~nm, and we used the OG515\_G0330 filter to block light blueward to 515~nm. The blocking filter was used to avoid contamination from light from higher orders. For all observations, we windowed the regions of interest (ROI) on the detector in the cross-dispersion direction to reduce the readout time and improve the duty cycle. We used one ROI for each slit, covering the whole detector in the dispersion direction and approximately 40~arcseconds in the cross-dispersion direction. 

Each observation covered the transit of the planet (lasting 1.58 hours) and additional out of transit baseline, and in total lasted approximately 3.5 hours. One of the comparison stars we observed is 2MASS J09534228-4538376 (G$_{mag}$ = 13.52, (\citealt{Gaia2018}), hereafter referred to as comparison star 1) at a distance of $\sim 1$~arcmin from WASP-19 (which has G$_{mag}$ = 12.1)  and 1.22 magnitudes fainter than WASP 19. We also observed an additional brighter star TYC 8181-2204-1 ($\sim 2.12$~arcmin away and G$_{mag}$ = 11.14, (\citealt{Gaia2018}), hereafter referred to as comparison star 2) simultaneously with WASP-19 and J09534228-4538376. However, the longer exposure times required in order to improve the duty cycle and gain adequate signal-to-noise for WASP-19 meant that the brighter comparison star is saturated in some exposures. In addition, there is a large group of bad columns in the location of the sodium feature in the stellar spectrum of the bright star, which could not be avoided without significantly altering the telescope PA. Hence, we choose to use only comparison star 1 for all further analysis in this paper.

In order to avoid slit losses, we chose to use MOS masks with wide slits of 10~arcsec width for each star. The slits were kept 30~arcsec long to ensure adequate background sampling for each star. In order to make sure that the spectra of both stars had a similar wavelength coverage for each observation, we selected the PA of the MOS mask to be as close as possible to the PA between WASP-19 and comparison star 1 (202.7 degrees E of N).

The GMOS-S detector was replaced in June 2014, during our survey program, to reduce the effects of fringing and improve sensitivity in the red optical \footnote{https://www.gemini.edu/sciops/instruments/gmos/\\imaging/detector-array/gmosn-array-hamamatsu?q=node/10004}. As a result, transits~1 through~7 were obtained using the original detector, manufactured by e2v, while transit~8 was obtained using the new detector, manufactured by Hamamatsu (\citealt{Scharwachter2018}). Three amplifiers were used for R150 transits 1 through 7, and we used the $1\times2$ readout mode to reduce overheads, binning only in the cross-dispersion direction. The amplifier gains for transit 1 to 7 varied from 1.63 to 1.52 ~$e^-$/ADU. For transit~8, the new setup used 12 amplifiers simultaneously, which reduced overheads enough that we were able to use the $1\times1$ binning mode. The amplifier gains for transit~8 varied from 1.61 to 1.85~$e^-$/ADU. Exposure times for all observations were chosen to keep the count levels between 10,000 and 40,000 peak~ADU and well within the linear regime of the CCDs. Table \ref{obsstats} gives more details on the observation log for each transit. The numbers given under `No.' in Table \ref{obsstats}  are the numbers by which we will refer to each transit observation in this paper.

\begin{table*}
\centering
\begin{tabular}{ccccccccc}
\hline
\hline
No. & Observation ID & UT Date & Exposure & No. of & Duty & Seeing & Airmass \\
& & & Time (s) & Exposures & Cycle (\%) &  (arcsec) & Range \\
\hline
1 & GS-2012B-Q-41 & 2013 Jan 24   & 80 & 108 & 71 & 1.17 & 1.04-1.22  \\
2 & GS-2012B-Q-41 & 2013 Feb 4    & 33-65 & 173 & 57 & 0.87-0.82   & 1.04 - 1.22 \\
3 & GS-2012B-Q-41 & 2013 Feb 12   & 47-65 & 140 & 61 &  0.55-1.11 & 1.04-1.23 \\
4 & GS-2013B-Q-44 & 2014 Jan 10   & 50 & 153 & 59 & 0.67-0.92  & 1.04-1.24 \\
5 & GS-2014A-Q-32 & 2014 Feb 9    & 60 & 137 & 62 &  0.34-0.91 & 1.04-1.14 \\
6 & GS-2014A-Q-32 & 2014 Mar 11   & 68 & 120 & 66 & 0.66-0.78  & 1.04-1.42 \\
7 & GS-2014A-Q-32 & 2014 Apr 10   & 80 & 108 & 70 & 0.65-0.95  & 1.10-1.95 \\
8 & GS-2014B-Q-45 & 2014 Dec 31   & 65 & 123 & 57 & 0.76-1.05 & 1.05-1.68 &  \\
\hline
\end{tabular}
\caption[1]{Observing Conditions for transits of WASP-19b at GMOS-South. The numbers in the first column are the numbers by which we will refer to each transit throughout the rest of the paper. Note that the seeing was worse during observation 1 and exposure times were varied frequently throughout observation 2 and 3.}
\label{obsstats}
\end{table*} 


\section{Data Reduction}
\label{sec_data_reduc}


\subsection{Data reduction of the Gemini/GMOS observations}
\label{gmos_data_reduc}

We used our custom pipeline designed for reducing the GMOS data, the steps for which are described in more detail in \citetalias{Huitson2017}. We extract the 1D spectra and apply corrections for additional time- and wavelength-dependent shifts in the spectral trace of target and comparison stars on the detector due to atmospheric dispersion and airmass. In this section, we describe the main steps of the pipeline and the additional corrections we apply to the data before extracting and analysing the transit light curves.


\subsection{Flat-Fielding}

We acquired 100 flat frames for transits 1 through 7 and 200 flat frames for transit 8 with median count levels of $\sim 10,000$ ADU. For transits 1 through 7, both the flat field and science frames show fringing at the 10~\% amplitude. We construct a master flat by median combining the series of flats for each observation. We found that the scatter in the transit light curves redward of 700~nm was 10~$\times$ photon noise without flat-fielding, which is marginally higher when performing flat-fielding. On inspection of the frames, we found that noise is added by flat-fielding because the phase, period and amplitude of the fringe pattern are significantly different between the flat fields and the science frames. The fringe pattern in the science frames also changes by several times the photon noise during each transit observation. For transit~8, the fringe amplitude is an order of magnitude lower than in the other transits. However, flat-fielding still increased the scatter redward of 700~nm by 10-20~\%. We attribute this to low-levels of fringing still being present in the transit 8 observations taken with the new detector. Moreover, flat-fielding should not be a major issue since we measure the transit depth for the same set of pixels relative to time. However, changes in the gravity vector of the instrument due to changing pointing through the night can cause the spectral trace to drift to different sets of detector pixels during the observation. We tested our extraction with and without flat-fielding and find that flat-fielding does not significantly affect the scatter of the resulting transit light curves. We found that the flat fielding changed the light curve scatter on average by 40 ppm across all 8 transit light curves, which is about 10 times lower than the typical photon noise for a GMOS transit light curve of WASP-19b. For this reason, and since flat-fielding did not improve the scatter blueward of 700~nm, we chose not to perform flat-fielding for all transit observations. We notice no slit tilt in the spectra of WASP-19 and the comparison stars, unlike as seen in \citetalias{Huitson2017} and \cite{Todorov2019}. The skylines in the frames for all transits are parallel to the pixel columns. Thus, we choose to not perform any tilt correction. 


\subsection{Spectral extraction}

We follow the steps outlined in more detail in \citetalias{Huitson2017} to detect and mask cosmic ray hits and bad pixel columns (mainly due to shifted charge) from all science frames. We then subtracted the background while performing the optimal extraction (\citealt{Horne1986}), and found that subtracting the median value in each cross-dispersion column provided the best fit to the background fluxes compared with performing fits to the flux profile in each cross-dispersion column. We also found that the precision in the light curves was $\sim 2 \times$ better when using the median value for each column rather than a fit.

To test the degree to which background subtraction affects our resultant transmission spectra, we also extracted spectra in which the background subtraction was multiplied by 10$\times$. We found that all $R_\mathrm{P}/R_\star$ values in the final transmission spectra deviated by much less than 1-$\sigma$ between the two cases, indicating that the final results are robust to uncertainties in background subtraction. The background flux level was 1.4-10~\% of the stellar flux for WASP-19, 5-25~\% of the stellar flux for comparison star 1 and 0.5-3~\% of the stellar flux for the comparison star 2. 

We additionally also extract the average PSF width of the spectral trace, which we use later for transit light curve modelling in Section \ref{sec:wtlc}. For each exposure, we first bin the 2D spectral trace in the raw science frames at an arbitrary interval of 10 pixels in the dispersion direction. To this binned spectral trace for each column, we then fit Gaussians along the cross-dispersion. We then take the average best fit FWHM of all the Gaussians to obtain the PSF width for each exposure.   

\subsection{Wavelength calibration}

After spectral extraction, we performed wavelength calibration using CuAr lamp spectra taken on the same day as each science observation, following similar steps as described in \citetalias{Huitson2017} and \citetalias{Panwar2022}. The final uncertainties in the estimated wavelength solution are approximately 1 nm for all observations, which is $\sim5$\% of the size of wavelength bins (20 nm) we use in the final transmission spectrum for all transits in Section \ref{sec:binned_LC}. This level of uncertainty in the wavelength solution is smaller than our resolution element ($\sim4$\,nm in R150 e2v detectors) and is insufficient to cause systematic effects in the final wavelength-dependent light curves.

\begin{figure}
  \centering
  \includegraphics[scale=0.6]{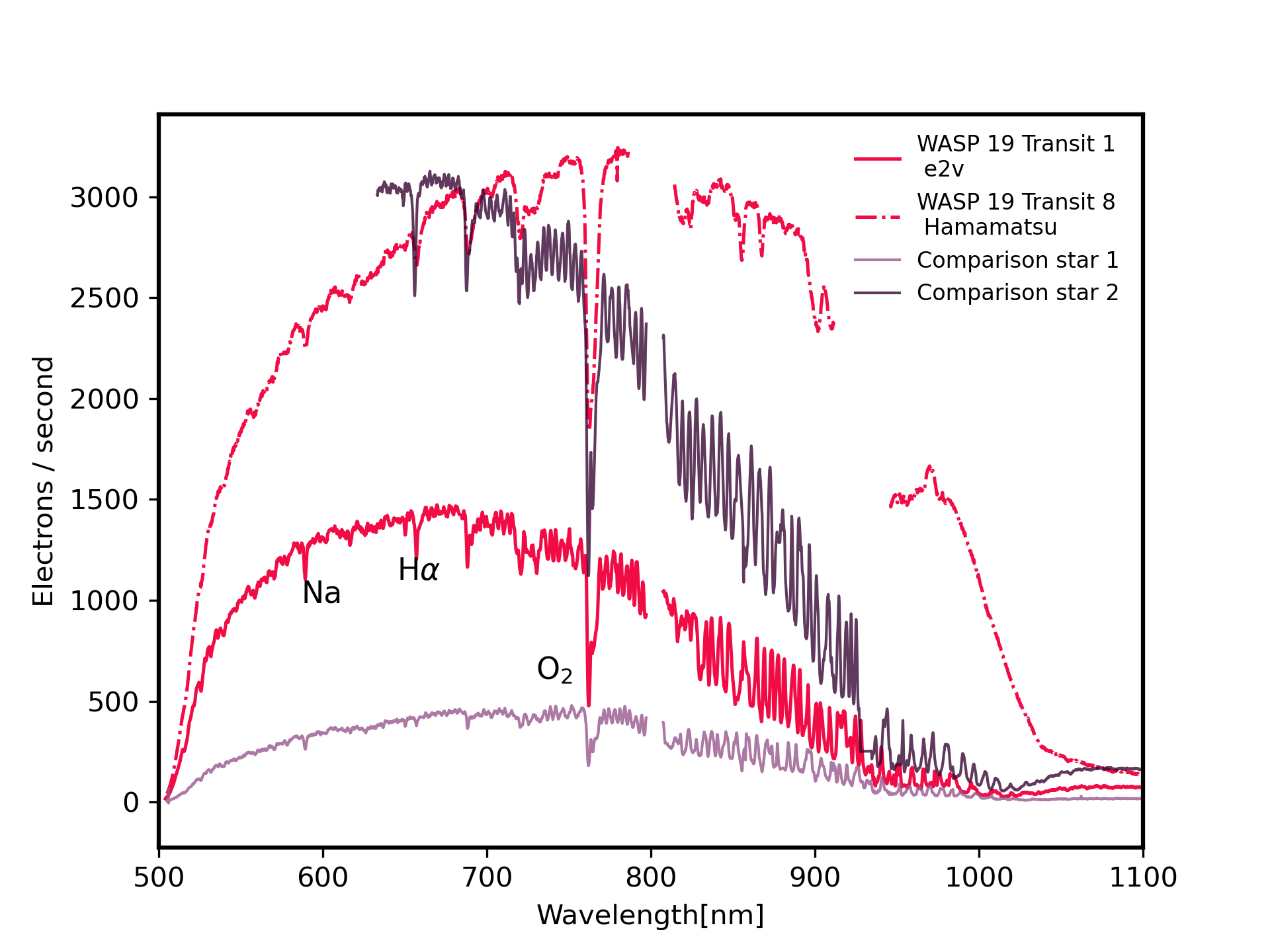}
  \caption[1]{ One of the extracted 1D spectra for target and comparison stars for GMOS-R150 observation 1 and observation 8 of WASP-19b. The two comparison star spectra shown here are both for observation 1 to illustrate the relative brightness difference between the target and comparison stars. All spectra were extracted at a similar airmass and normalized by the exposure time. Prominent stellar and telluric features which we use for measuring the shifts in spectral trace with time are labelled. It can be seen that the fringing gets significant at wavelengths longer than 720 nm for observation 1 (and similarly for observations 2 to 7) but is reduced in observation 8 taken with the new detector. The gaps in wavelength coverage are due to the physical gaps between individual CCDs in the detector, and deviation from the PA of the telescope for comparison star 2.}
  \label{fig:1D_spectra}
\end{figure}

\subsubsection{Dispersion-direction shifts of the stellar spectra}

The wavelength solution for GMOS data is known to shift and stretch with time because of the absence of atmospheric dispersion compensator (ADC). These shifts and stretches vary both in time and wavelength and manifest as a slope in the measured transmission spectrum of the planet if not corrected for, as demonstrated by \citetalias{Huitson2017}. In a recent study, \cite{Pearson2018} introduced a method to measure Gemini/GMOS spectral shifts by computing the cross power spectrum of the stellar spectra in the Fourier space, also known as phase-only correlation algorithm. This is equivalent to performing cross-correlation of the stellar spectra in the wavelength space, as done by \citetalias{Huitson2017}. We follow the \citetalias{Huitson2017} approach which is also described in detail in \citetalias{Panwar2022}, and select three features (Na, H$\alpha$, and O$_{2}$) in the stellar spectrum of WASP-19 (also labelled in Figure \ref{fig:1D_spectra}). In brief, we measure the spectral shifts for each feature in time by cross-correlation of each of the 1D spectra with a reference spectrum obtained around the mid-transit. The measured spectral shifts are then used to apply interpolated corrective shifts to every pixel for each exposure. We repeat this step for both the comparison stars as well, using the same set of spectral features as the target star spectrum. To correct for shifts between the target and comparison stars themselves, we then interpolate the comparison stars' spectra for each exposure onto the interpolated common wavelength solution of the target star, omitting detector gaps and bad columns. This results in a common wavelength solution for both the target and comparison star spectra. We apply these corrections to each observation.

However, for all observations we find that the transmission spectrum we obtain from the GP based methods we use in \citetalias{Panwar2022} are consistent within 1$\sigma$ whether we perform the spectral shift and stretch corrections or not. This indicates that the GP model from \citetalias{Panwar2022} which we eventually use to fit the spectroscopic light curves (described in more detail in Section \ref{sec:binned_LC}) mitigates the effects of stellar spectral shifts and stretch on the final transmission spectrum. Hence, we opt to use the optimally extracted spectra without any shift and stretch corrections. Moreover, we eventually use only the target spectroscopic light curves to extract the transmission spectrum, which prevents the effects of shifts between the target and comparison star spectra. We additionally also use a wavelength bin size of 20 nm, which is significantly larger than the average amplitude of spectral shifts.    

\subsection{Extracting the light curves}
\label{sec:construct_tlc}

After extracting the time series of the 1D spectra for the target and comparison stars for each transit observation, we proceed to construct the corresponding light curves. We construct the white light curves for both the target and comparison stars by summing the flux for each exposure spectrally over the wavelength range of 520 to 720 nm for transit 1 to transit 7, and from 520 to 900 nm for transit 8. Since the exposure time in general was not fixed throughout the night, we also normalized the total flux in each exposure by the corresponding exposure time. We then normalize the comparison star light curve by its median, and the target star light curve by the median of the out of transit exposures. For constructing the spectroscopic light curves, we repeat the same process for each of the 20 nm wide wavelength bins.          


\section{Analysis}
\label{sec_analysis}

We now describe our light curve analysis as applied to the 8 GMOS transit observations of WASP-19b with the goal to obtain the planet's transmission spectrum. We first discuss the analysis of the white transit light curves in Section~\ref{sec:wtlc} for which we use two independent methods: the conventional method that fits for the Target/Comparison light curves, and the new method recently introduced by \citetalias{Panwar2022} of fitting the target star light curves directly using the comparison star light curve as one of the GP regressors. In Section~\ref{sec:binned_LC} we describe the analysis of the wavelength binned light curves also using the conventional method and the new method from \citetalias{Panwar2022} to obtain the transmission spectrum for each observation. 

\subsection{Modelling systematics in GMOS transit light curves}
\label{sec:noise_model}

We model the instrumental and telluric time-dependent systematics in the WASP-19 transit light curves constructed in Section \ref{sec:construct_tlc} following the conventional as well as the new method introduced and described in more detail in \citetalias{Panwar2022}. The conventional method involves a linear approach of first normalizing the target light curve by the comparison star light curve to correct for systematics commonly affecting both the target and comparison star light curves. The resulting Target/Comparison light curve typically still suffers from residuals systematics arising from non-linear differences at the telluric level e.g. brightness or spectral type between the target and comparison stars (leading to different telluric systematics in their respective light curves). This differences can also arise at the instrument level e.g. due to a non-ideal PA, unequal travel times of the instrument shutter common in GMOS observations. The Target/Comparison light curve is then fit with a transit model added to a parametric or a non-parametric (e.g. Gaussian Processes (GP) \citealt{Gibson2012}) systematics model. 

The new method introduced by \citetalias{Panwar2022} (\texttt{New:WLC} followed by \texttt{New:$\lambda$LC}, as described in Table 2 in \citetalias{Panwar2022}) fits the target transit light curve directly, using the comparison star light curve as one of the regressors in a GP systematics model. The advantage of the new method is that it avoids introducing unwanted systematics to the target light curve as a result of normalization by a non-ideal comparison star light curve that behaves differently during the night. The new method in fact allows using the comparison star light curve as one of the regressors to the GP model (for the white target light curves) and let the GP glean the likely non-linear mapping between the systematics common to both the target and comparison light curves. In this process, it also propagates the uncertainties within a Bayesian framework instead of simple addition by quadrature (as is the case when doing Target/Comparison normalization). This approach is further relevant to our observations of WASP-19 as the only comparison star we have at our disposal is significantly fainter ($\sim$ 1.22 magnitude fainter) as compared to WASP-19. Moreover, we are already dealing with a host star whose stellar variability has a significant effect on the transmission spectra (as discussed in Section \ref{sec:stellar_activity_discuss}. Additional stellar variability of the comparison star can lead to further contamination of the final transmission spectrum due to wavelength dependent effect present in the comparison star spectroscopic light curves themselves. For the comparison stars observed using Gemini/GMOS we do observe stellar variability in their TESS light curves albeit at lower amplitudes as compared to WASP-19 (described in more detail in Appendix \ref{tess_lc_analysis_comp_stars}). Hence, it is important to not directly use the comparison star spectroscopic light curves when measuring the final transmission spectrum. Our new method only uses the comparison star white light curve to fit the target star white light curve and then uses the target star common-mode trend to fit the spectroscopic target light curves, as we describe in more detail in Section \ref{sec:binned_LC}.         

Both the methods have the common aspect of fitting the transit light curve as a systematics model added to a numerical transit model. The main difference between the two methods is that the new method uses the GP framework of \cite{Gibson2012} to model the systematics directly in the target star light curves, accounting for the non-linear differences between the target and comparison star light curves. In this method, the comparison star light curves are essentially used as a control sample to check that the noise is efficiently modelled. 

The GP model we use for modelling the systematics for both methods (i.e., for modelling both the Target/Comparison and Target star light curve respectively) is the same as that described in more detail in \citetalias{Panwar2022}. In brief, we use a Mat\'ern 3/2 kernel function to construct the GP covariance matrix, with a single amplitude hyperparameter and a length scale hyperparameter for each of the inputs to the GP.  

\subsection{Analysis of White Transit Light Curves}
\label{sec:wtlc}

\begin{figure*}

  \centering
  \includegraphics[scale=0.4]{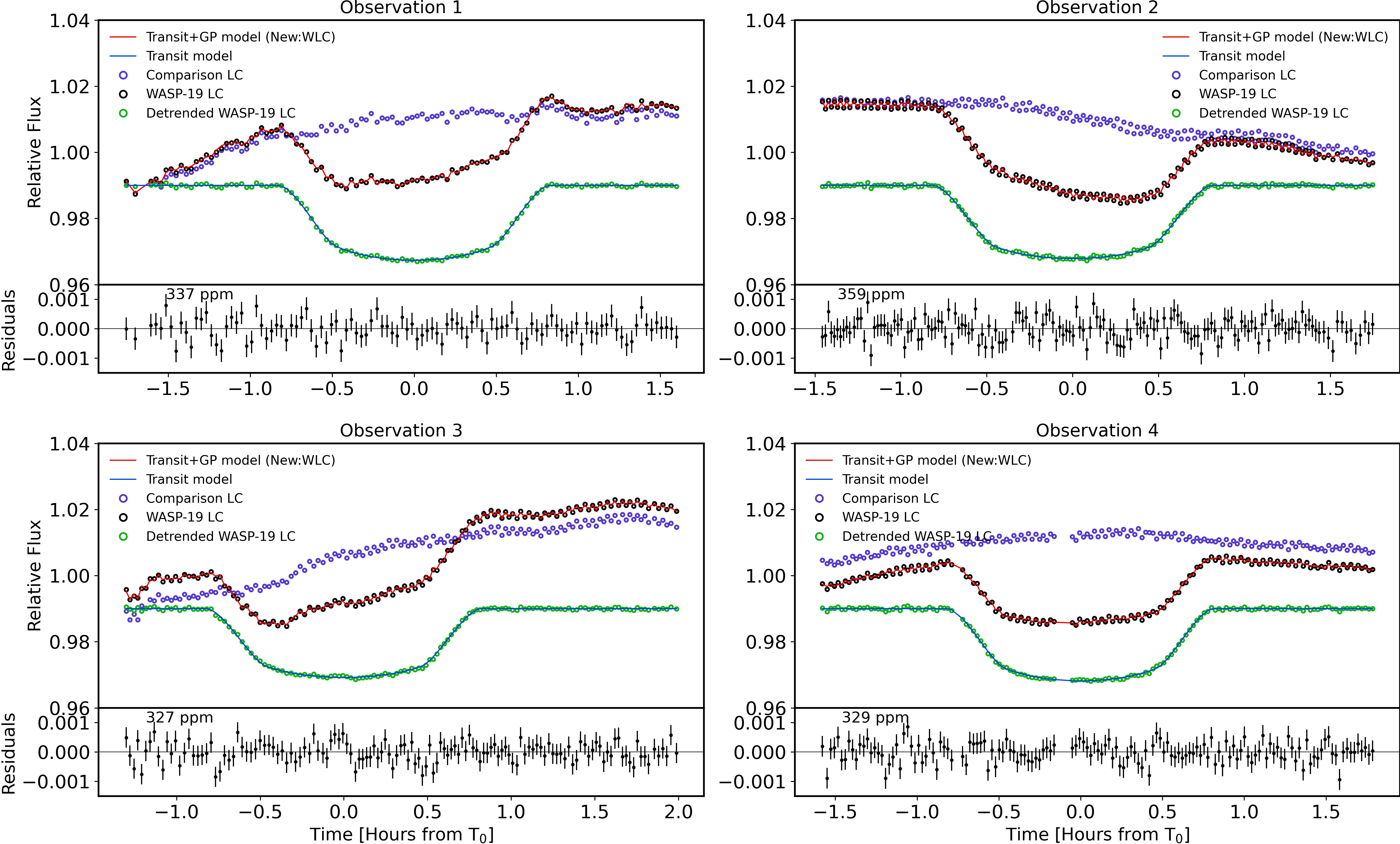}
  \caption[1]{White GMOS-R150 target light curves for WASP-19b and their best fits obtained using the new method from observations 1 to 4. For all observations the black points are the target light curves overplotted with the best fit transit + GP systematics model in red, purple points show the comparison star light curve, and green points are the detrended target light curve overplotted with the best transit model in blue. The detrended light curve and the respective best fit transit model have been offset for clarity. Note that for all observations we observe significant odd-even effect in both the target and comparison star light curves, which are efficiently modelled by the GP model in the new method using the comparison star as one of the GP regressors. The gap in the light curve for observation 4 is due to outliers in the light curve around the inflection point for the Cassegrain rotator which happens when the target reaches zenith. 
  }
  \label{fig:wtlc_1to4}
\end{figure*}

\begin{figure*}

  \centering
  \includegraphics[scale=0.4]{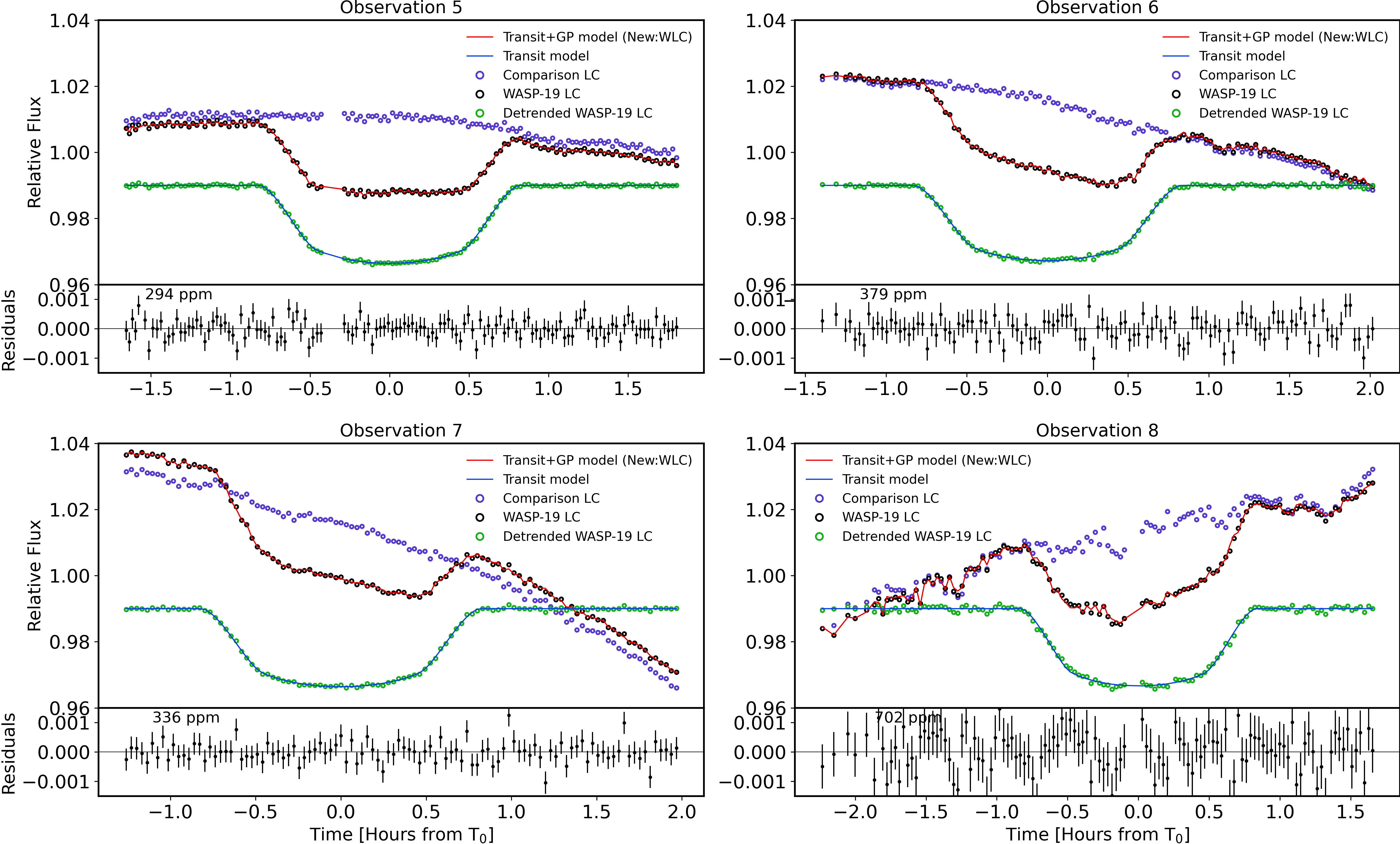}
  \caption[1]{Same as Figure \ref{fig:wtlc_1to4} but for observations 5 to 8. Note that observation 8 is noisier as compared to other observations because it was taken using a different detector (as described in \ref{gmos_obs}) and setup as compared to all the other observations. The gaps in the light curves for observations 5 and 8 are due to outliers in the light curve around the inflection point for the Cassegrain rotator which happens when the target reaches zenith.    
  }
  \label{fig:wtlc_5to8}
\end{figure*}

We describe some steps and details common to both methods (i.e., both Target/Comparison and Target light curves ) mentioned in Section \ref{sec:noise_model} as applied to the transit white light curves for all the 8 transits, and mention specifically the points at which the two methods differ. 

We use the transit modelling package {\texttt{batman}} (\citealt{Kreidberg2015}) to calculate the numerical transit model $\mathbf T(\bmath t,\phi)$ (where \textbf{t} are the time stamps of each exposure and  \textbf{$\phi$} is the set of orbital transit parameters), and the package {\texttt{george}} (\citealt{Ambikasaran2015}) for constructing and computing the GP kernels and likelihoods. In Table \ref{tab:priors} we summarize the parameters we fix and the priors we employ in our fitting procedure. We fix the orbital period ($P$) and eccentricity ($e$), and fit for the orbital inclination ($i$), orbital separation ($a/R_\star$), central transit time ($T_0$), planet to star radius ratio ($R_\mathrm{P}/R_\star$). We employ a linear limb darkening law and fit for the linear limb darkening coefficient u$_{1}$. We choose to use linear limb darkening law because given the precision and time resolution of our light curves, multiple free parameters describing the limb darkening e.g. in case of quadratic or non-linear limb darkening law, are difficult to constrain. Hence, the linear law in this context is the simplest choice to fit for. Recent work by \cite{PatelEspinoza2022} demonstrates that specifically quadratic limb darkening coefficients for sun-like or cooler stars often suffer from discrepancies between the theoretical and empirical methods used to estimate them. 

For each transit model parameter except the linear limb darkening coefficient we put truncated uniform priors within $\sim$10 $\sigma$ bounds around their literature values. We calculate the linear limb darkening coefficient u$_{1}$ for the wavelength range integrated to obtain the white light curve, and the wavelength bins we adopt for spectroscopic light curves (see Section \ref{sec:binned_LC}) using \texttt{PyLDTk} (\citealt{Parviainen2015}), which uses the spectral library in (\citealt{Husser2013}), based on the stellar atmosphere modelling code PHOENIX. We put a Gaussian prior on the linear limb darkening coefficient with the mean value and the standard deviation as the mean and 3 times the 1 $\sigma$ uncertainty calculated from \texttt{PyLDTk} respectively. We also fit for the white noise parameter $\sigma_{w}$ which lets the GP model fit for the white noise variance in the target star light curves and also includes contribution from the variance inherent in a noisy GP input itself (e.g. the comparison star light curve). This is in fact an important feature of the new method and provides a natural way to propagate uncertainties from the comparison star light curve to our fit of the target star light curve within a Bayesian framework. We emphasize that fitting for $\sigma_{w}$ is crucial for letting the GP model capture the white noise in the target star light curves. 

We perform the white light curve fits using all possible combinations of GP input parameters used to construct a combined Mat\'ern 3/2 kernel function (described in more detail in \citetalias{Panwar2022}). For the conventional method, we use time, CRPA (Cassegrain Rotator Position Angle), and airmass as GP regressors. For the new method, we use the same set of GP regressors as the conventional method but additionally also the point spread function (PSF) width of the target spectral trace for every exposure, and the comparison star light curve.

We put wide uniform priors on the logarithm of the GP hyperparameters which include the covariance kernel function amplitude $A$ and the length scales for each GP regressor $\eta_{p}$. We effectively sample the amplitude and length scale hyperparameters logarithmically as shown in Table \ref{tab:priors}. The logarithmic sampling of hyperparameters effectively puts a shrinkage prior on them, which pushes them to smaller values if the corresponding input vector truly does not represent the covariance in the time series (\citealt{Gibson2012}). 

\begin{table}
\centering
\caption[1]{Summary of priors and fixed values for the parameters (\texttt{batman} transit model and GP hyperparameters) used to fit the transit light curves of WASP-19b. For all the fits we fixed the planet orbital period (P) and eccentricity (e). $\mathcal{U}$ shows a uniform prior applied within the specified range, and $\mathcal{N}$ represents a Gaussian prior with the mean and standard deviation respectively. T$_{c}$ is the predicted mid-transit time for each epoch using the ephemeris from \citealt{Hartman2011}. For the limb darkening we use a Gaussian prior around the mean linear limb darkening coefficient theoretically calculated by \texttt{PyLDTk} (\citealt{Parviainen2015}) corresponding to the stellar parameters in Table \ref{tab:stellar_params} and for the R150 wavelength range of 520 to 900 nm.}

\begin{tabular}{ccc}
\hline
\texttt{batman} model parameters \\
\hline
\hline
Parameter & Prior/Fixed value & Reference \\
\hline
P [d] &  0.7888390 & \cite{Lendl2013} \\
$e$ & 	0.0046  &  \cite{Hellier2011} \\
$i$ [$^\circ$] & $\mathcal{N}$ (79.5, 1.5)  & \cite{Lendl2013}\\
$R_{p}/R_{*}$ & $\mathcal{U}$ (0, 1)   & -- \\
$a/R_{*}$ & $\mathcal{N}$ (3.573, 1.5) & \cite{Lendl2013} \\
$T_{0}$[d] & $\mathcal{U}$ (T$_{c}$-0.001, T$_{c}$+0.001) & \cite{Hellier2011} \\
u$_{1}$[R150] & $\mathcal{N}$ (0.63, 0.03)  & \texttt{PyLDTk}  \\  \\ \\
\hline
GP model hyperparameters \\
\hline
\hline
ln (A) & $\mathcal{U}$ (-100, 100)  & -- \\
ln ($\eta_{p}$) & $\mathcal{U}$ (-100, 100) & -- \\
$\sigma_{w}$ &  $\mathcal{U}$ (0.00001, 0.005) & -- \\
\hline
\end{tabular}

\label{tab:priors}
\end{table} 

We first find the Maximum a-Posteriori (MAP) solution by optimizing the GP posterior (see \citetalias{Panwar2022} and \citealt{Gibson2012} for more detail) using the Powell optimizer in the {\texttt{SciPy}} python package. 

Using the MAP solution as the starting point, we marginalize the GP posterior over all hyperparameters and transit model parameters through an MCMC using the package {\texttt{emcee}} (\citealt{Goodman2010}, \citealt{ForemanMackey2013}). We use 50 walkers for 10000 steps and check for the convergence of chains by using the integrated autocorrelation times for each \texttt{emcee} walker following the method described in (\citealt{Goodman2010}). We ensure that the total length of our chains is greater than 50 times the integrated autocorrelation time which indicates that our samples are effectively independent and have converged.  We also tested the robustness of our posteriors from a nested sampler using the package {\texttt{dynesty}} (\citealt{Speagle2020}) and obtain posteriors consistent with those from {\texttt{emcee}} well within 1 $\sigma$. We estimate the best fit parameters by taking the 50th percentile and their $+$1 $\sigma$ and $-$1 $\sigma$ uncertainties by taking the 84th and 16th percentile respectively of the MCMC posteriors. We show and compare the best fit transit parameters (corresponding to the combination of GP inputs that perform best for both methods) and their $\pm$ 1 $\sigma$ uncertainties in Table \ref{tab:bestfitparams}. In Figures \ref{fig:wtlc_1to4} and \ref{fig:wtlc_5to8} we show the best fits to the target star light curve obtained from the new method for all 8 observations. 

We select the best GP regressor combination for both methods independently using two criteria: 1) Bayesian evidence (log$_{e}$Z) estimate from \texttt{dynesty} and 2) the Bayesian Information Criterion (BIC, \citealt{Schwarz1978}) computed using the GP likelihood corresponding to the best fit transit model parameters and hyperparameters. We use the $\Delta$BIC and $\Delta$log$_{e}$Z threshold prescribed by \cite{KassRaftery1995} and also used in \citetalias{Panwar2022} to choose the best GP regressor combination for the two methods individually. We find that the GP regressor combination selection based on both the criteria (BIC and log$_{e}$Z) always agree within their model selection thresholds as prescribed in \cite{KassRaftery1995}. We show the best GP regressor combination and the best fit transit parameters and their 1$\sigma$ uncertainties for both the conventional and new methods in Table \ref{tab:bestfitparams}. The best fit GP hyperparameters and their uncertainties are shown in Table 1 in the supplementary material. In Table \ref{tab:litfitparams} we list the transit parameters measured by previous studies and this work using TESS photometry as described in Section \ref{tess_lc_analysis}.        

For the new method, using the comparison star as one of the GP regressors gives the best fit for most of the observations. Specifically, for the new method applied to observations 3 and 6, we find that using only the comparison star light curve as the GP regressor performs best. For all other observations, using time or airmass as a regressor in addition to the comparison star light curve helps to model the lower frequency variations in the target star light curve which are not present in the comparison star light curve.

For the conventional method, using just time as a GP regressor gives the best fit for most observations. We find that the new method for most of the observations, and in particular observation 8, gives comparable or better fits compared to the conventional method when considering the transit depth precisions and the residual RMS. The new method yields on an average 10 to 20 \% smaller RMS on the residual scatter for the best fit as compared to the conventional method.  

\subsubsection{Correcting for the odd-even effect in the light curves}
\label{sec:odd_even} 

The consecutive exposures in the GMOS light curves suffer from an odd-even effect due to unequal travel times of the GMOS blade-shutters with respect to the direction of motion, and have been previously observed and corrected for in \citetalias{Panwar2022} and \cite{Stevenson2014}. We estimate the level of this effect for our WASP-19b observations to be around $\sim$300 ppm for both the target and comparison star light curves. This effect is most significantly observed in observations 2, 3, 4, 5, and 6 (Figures \ref{fig:wtlc_1to4} and \ref{fig:wtlc_5to8}). Note that the amplitude of this odd-even effect is not exactly the same for both the target and comparison star light curves (since it depends on the direction of motion of blade-shutters). This difference in the amplitudes was in fact observed for one of the transits of HAT-P-26b in \citetalias{Panwar2022} (labelled as observation 2 in that paper). It was observed that due to the difference in the timing of this odd-even effect for the target and comparison light curves respectively, simply dividing the target by the comparison light curves as done during the conventional method doesn't correct for this effect and instead exacerbates it. This is one of the examples of a non-linear relationship between how the same source of systematics affect the target and comparison star light curves. The new method resolves this by letting the GP determine this non-linear mapping. Note that the timescale of the odd-even effect is the same for both target and comparison star light curves as we confirm from their individual Lomb Scargle periodograms. Using the comparison star light curve as a GP regressor as in the new method is able to efficiently model this effect, as can be observed in the best fit models and the residuals in Figures \ref{fig:wtlc_1to4} and \ref{fig:wtlc_5to8}.          

Once we retrieve the best fit transit parameters for each observation from the respective white transit light curve, we use this information to fit the spectroscopic light curves and obtain the transmission spectrum as described in more detail in Section \ref{sec:binned_LC}. 


\begin{table*}
\begin{center}

\caption[1]{Best fit transit parameters obtained from the fits to white transit light curves of eight GMOS-R150 observations analysed in this work. Two rows for each observation as specified in the second column compare the best fit transit parameters and residual RMS resulting from the new method from \citetalias{Panwar2022} of fitting the Target white light curve and the conventional method of fitting the Target divided by the Comparison light described in more detail in Section \ref{sec:noise_model}. The third column specifies the best GP regressor combination for the GP model for both methods for each observation, as described in more detail in Section \ref{sec:wtlc}. `Time' refers to the time stamps of the exposures in the observation, and `Comp' refers to the comparison star light curve. $\sigma_w$ values are the median white noise value quantifying the diagonal of the GP covariance matrix with a measured uncertainty. The RMS values are the standard deviation of the residuals between the light curve and the transit model and the predicted mean of the GP systematics model corresponding to the median of the posteriors. Both of these quantities can be viewed as two ways of estimating the white noise level of the light curves.} 
\label{tab:bestfitparams}

\begin{tabular}{ccccccccccc}
\hline
\hline 
No.& Method & GP regressors & $R_{p}/R_{*}$ & $T_{0}$[$BJD_{TDB}]$ & $a/R_{*}$ & $i$ [$^\circ$] & u$_{1}$ & $\sigma_{w}$ [ppm] & RMS [ppm]   \\
\hline
1 & New & Time, Comp & 0.1449$^{+0.0037}_{-0.0035}$ & 2456316.730224$^{+0.000194}_{-0.000184}$ & 3.6$^{+0.05}_{-0.05}$ & 79.88$^{+0.42}_{-0.41}$ & 0.63$^{+0.03}_{-0.03}$ & 383$^{+35}_{-28}$ & 337 \\
  & Conventional & Airmass & 0.1389$^{+0.0012}_{-0.0012}$ & 2456316.729882$^{+0.000237}_{-0.000237}$ & 3.52$^{+0.07}_{-0.06}$ & 79.11$^{+0.51}_{-0.49}$ & 0.64$^{+0.02}_{-0.03}$ & 395$^{+39}_{-30}$ & 349 \\ \\

2 & New & Time, Comp & 0.1451$^{+0.0009}_{-0.0009}$ & 2456327.773539$^{+7.4e-05}_{-7.3e-05}$ & 3.57$^{+0.03}_{-0.03}$ & 79.23$^{+0.22}_{-0.21}$ & 0.59$^{+0.02}_{-0.02}$ & 373$^{+25}_{-25}$ & 359  \\
  & Conventional & Time & 0.1451$^{+0.0012}_{-0.0012}$ & 2456327.773395$^{+0.00012}_{-0.000101}$ & 3.58$^{+0.04}_{-0.03}$ & 79.29$^{+0.26}_{-0.25}$ & 0.61$^{+0.02}_{-0.02}$ & 431$^{+27}_{-23}$ & 416  \\ \\

3 & New & Comp & 0.1408$^{+0.0009}_{-0.0009}$ & 2456335.662255$^{+8.1e-05}_{-8.2e-05}$ & 3.52$^{+0.04}_{-0.03}$ & 79.1$^{+0.28}_{-0.24}$ & 0.58$^{+0.02}_{-0.02}$ & 345$^{+24}_{-21}$ & 327 \\
  & Conventional & Time& 0.1398$^{+0.0022}_{-0.0023}$ & 2456335.662113$^{+0.000124}_{-0.000127}$ & 3.53$^{+0.04}_{-0.04}$ & 79.16$^{+0.3}_{-0.27}$ & 0.61$^{+0.03}_{-0.02}$ & 338$^{+24}_{-22}$ & 316 \\ \\

4 & New & Time, Comp, Airmass & 0.1434$^{+0.0021}_{-0.0021}$ & 2456667.763054$^{+0.000116}_{-0.000113}$ & 3.59$^{+0.04}_{-0.04}$ & 79.5$^{+0.3}_{-0.3}$ & 0.61$^{+0.03}_{-0.02}$ & 353$^{+25}_{-23}$ & 329 \\ 
  & Conventional & Time & 0.1391$^{+0.0034}_{-0.0029}$ & 2456667.763072$^{+0.000175}_{-0.000176}$ & 3.61$^{+0.06}_{-0.05}$ & 79.74$^{+0.43}_{-0.41}$ & 0.63$^{+0.02}_{-0.03}$ & 375$^{+25}_{-25}$ & 346 \\ \\

5 & New & Time, Comp & 0.1488$^{+0.0035}_{-0.0033}$ & 2456697.739132$^{+0.000155}_{-0.000157}$ & 3.6$^{+0.04}_{-0.04}$ & 79.64$^{+0.3}_{-0.29}$ & 0.6$^{+0.03}_{-0.03}$ & 326$^{+24}_{-22}$ & 294  \\
  & Conventional & Time, Airmass  & 0.1416$^{+0.0039}_{-0.0025}$ & 2456697.739098$^{+0.000201}_{-0.000216}$ & 3.54$^{+0.05}_{-0.05}$ & 79.17$^{+0.36}_{-0.41}$ & 0.6$^{+0.02}_{-0.02}$ & 380$^{+28}_{-26}$ & 346  \\ \\

6 & New & Comp & 0.1469$^{+0.0012}_{-0.0011}$ & 2456727.715017$^{+9.8e-05}_{-9.6e-05}$ & 3.52$^{+0.04}_{-0.04}$ & 79.05$^{+0.32}_{-0.32}$ & 0.62$^{+0.02}_{-0.02}$ & 402$^{+30}_{-25}$ & 379  \\
  & Conventional & Time & 0.1442$^{+0.0013}_{-0.0014}$ & 2456727.715003$^{+0.00011}_{-0.000123}$ & 3.52$^{+0.04}_{-0.04}$ & 79.05$^{+0.33}_{-0.35}$ & 0.62$^{+0.02}_{-0.03}$ & 399$^{+30}_{-25}$ & 378  \\ \\

7 & New & Time, Comp & 0.1487$^{+0.0032}_{-0.0035}$ & 2456757.690677$^{+0.000173}_{-0.000167}$ & 3.61$^{+0.05}_{-0.05}$ & 79.68$^{+0.34}_{-0.36}$ & 0.62$^{+0.03}_{-0.03}$ & 378$^{+39}_{-31}$ & 336   \\
  & Conventional & Time, Airmass & 0.1427$^{+0.0023}_{-0.0025}$ & 2456757.690571$^{+0.000165}_{-0.000212}$ & 3.59$^{+0.05}_{-0.05}$ & 79.57$^{+0.39}_{-0.35}$ & 0.63$^{+0.03}_{-0.01}$ & 402$^{+42}_{-36}$ & 356  \\ \\

8 & New & Time, Comp & 0.1482$^{+0.0033}_{-0.0038}$ & 2457022.740594$^{+0.000252}_{-0.000237}$ & 3.54$^{+0.03}_{-0.03}$ & 79.45$^{+0.24}_{-0.22}$ & 0.59$^{+0.01}_{-0.01}$ & 753$^{+65}_{-58}$ & 702  \\
  & Conventional & Time & 0.1371$^{+0.0059}_{-0.0054}$ & 2457022.740269$^{+0.000505}_{-0.000364}$ & 3.58$^{+0.05}_{-0.04}$ & 79.35$^{+0.25}_{-0.25}$ & 0.59$^{+0.01}_{-0.01}$ & 1051.0$^{+91}_{-81}$ & 958 \\ \\

\hline
\hline

\end{tabular}


\end{center}
\end{table*} 

\begin{table*}
\caption[1]{Transit parameters in the optical measured from Gemini/GMOS and other observatories (58 transits from TESS (analysed in this work), 1 transit from HST/STIS, 3 transits from VLT/FORS2,and 6 transits from Magellan/IMACS). Note that while for TESS we cite here the average values of measured transit parameters here, the $R_{p}/R_{*}$ values for TESS actually vary suggestively across the 58 transits (See Section \ref{sec:stellar_activity_discuss} and Figure \ref{fig:tess_tdepth_oot}). }
\label{tab:litfitparams}
\begin{tabular}{ccccc}
\hline
\hline
Instrument & $R_{p}/R_{*}$ & $a/R_{*}$ & $i$ [$^\circ$] & Reference \\
\hline
Gemini/GMOS & 0.1451$\pm$0.00051 & 3.57$\pm$0.013 & 79.45$\pm$ 0.099 & This work (8 transits) \\
TESS (600-1000 nm) & 0.1452$\pm$0.00035 & 3.58$\pm$0.015 & 79.78$\pm$ 0.099 & This work (TESS Sector 9 and 36) \\
HST/STIS (630-730 nm) & 0.1395 $\pm$ 0.0006 & 3.6 $\pm$ 0.5 & 79.8 $\pm$ 0.5 & \cite{Huitson2013} \\
VLT/FORS2 (400-1000 nm) & 0.14366 $\pm$ 0.00181 & 3.5875 $\pm$ 0.0574 & 79.52$^{+0.54}_{-0.56}$ & \cite{Sedaghati2017} \\
Magellan/IMACS (400-900 nm) & 0.14233 $\pm$ 0.0005 & 3.55 $\pm$ 0.014 & 79.29 $\pm$ 0.1 & \cite{Espinoza2019} \\

\hline
\hline 

\end{tabular}
\end{table*} 




\subsection{Analysis of Spectroscopic Light Curves}
\label{sec:binned_LC}

We now describe the analysis of the spectroscopic light curves (hereafter referred to as $\rm \lambda$LC) constructed by integrating the 1D stellar spectrum in 20 nm wide bins (as mentioned in Section \ref{sec:construct_tlc}. We chose the bin width of 20 nm as it is a few times the seeing limited resolution of $\sim$ 4 nm for our observations. This is similar to the previous R150 Gemini/GMOS observations from our survey program published by \citetalias{Huitson2017}, \cite{Todorov2019}, and \citetalias{Panwar2022}. For inspecting especially the bins centred around the 589 nm Na doublet we also construct spectroscopic light curves in 10 nm wide bins to sample the core and wings of the Na doublet.

We fit the spectroscopic light curves to extract the transmission spectrum using both the conventional method and the new method as introduced in \citetalias{Panwar2022} and described here in brief in the next two subsections. For both methods to fit the spectroscopic light curves, we follow the same procedure as the white light curves in Section \ref{sec:wtlc} to sample the posterior and obtain the best fit parameters and their uncertainties using \texttt{emcee} and \texttt{dynesty}.

\subsubsection{Conventional method using common-mode correction}
\label{sec:ts_old}

We first describe in brief the conventional method of fitting $\rm \lambda$LCs. We divide each target $\rm \lambda$LC by the corresponding comparison star $\rm \lambda$LC. GMOS $\rm \lambda$LCs are known to suffer from wavelength-independent systematics which are conventionally corrected for using common mode corrections (\citealt{Stevenson2014}, \citetalias{Huitson2017}, \citealt{Todorov2019}). We essentially use the GP noise model from the best fits to the Target/Comparison white light curves for each observation obtained in Section \ref{sec:wtlc} to do a conventional common-mode correction and remove time-dependent systematics common across all wavelength bins. 

For the conventional method, we derive the common-mode trend by subtracting the best fit white light curve transit model from the observed Target/Comparison white light curves. For each observation, this transit model is constructed using the corresponding best fit transit parameters obtained using the conventional method for the respective white Target/Comparison light curve as mentioned in Table \ref{tab:bestfitparams}. We then normalize the Target/Comparison $\rm \lambda$LC by their respective median out-of-transit flux and subtract the common-mode trend from each of them. We find that doing common-mode correction prior to fitting the Target/Comparison $\rm \lambda$LCs improved the precision of measured transit depths by $\sim15$\% on average per wavelength bin as compared to when we do not perform common mode correction. However, performing common-mode correction also implies that we effectively lose information on the absolute value of transit depths and the transmission spectra relative to the white light curve transit depth which was used to derive the common-mode trend.   

We fit the common-mode corrected Target/Comparison $\rm \lambda$LCs independently with the model described in \ref{sec:noise_model} as also used for white light curves in Section \ref{sec:wtlc}, using only time as a GP regressor. Using time as a GP regressor at this stage helps in accounting for residual wavelength-dependent systematics in the $\rm \lambda$LCs after common-mode correction, likely due to wavelength-dependent differential atmospheric extinction between the target and comparison stars with changing airmass through the night.

Since our main goal with the spectroscopic light curves is to measure the transit depth in each wavelength bin, we fix the orbital inclination ($i$), orbital separation ($a/R_\star$), and mid-transit time ($T_0$) to the best fit values for the corresponding white light curve in Section \ref{sec:wtlc} (see Table \ref{tab:bestfitparams}), and orbital period and eccentricity to literature values. We use a linear limb darkening law and employ a Gaussian prior for the limb darkening coefficients around the \texttt{PyLDTk} pre-calculated values for each wavelength bin (approximating a top hat transmission function for each wavelength bin). 

\subsubsection{New method using the common-mode trend as a GP regressor}
\label{sec:ts_new}

We now describe the application of the new method introduced in \citetalias{Panwar2022} to fit the Target $\rm \lambda$LCs directly. One of the motivations behind application of this approach is the large difference in brightness ($\sim$ 1.22 mag in V$_{mag}$) between the target and comparison star, which makes the correction for time and wavelength dependent systematics through differential spectrophotometry suboptimal and a source of additional uncertainties. Instead of dividing the target star $\rm \lambda$LCs by the comparison star $\rm \lambda$LCs and performing the conventional common-mode correction, we use the common-mode trend derived from the white target light curve as a GP regressor to fit the systematics and transit depth in the target $\rm \lambda$LC. We do this by first deriving the common-mode trend in the same way as done for the conventional method in Section \ref{sec:ts_old}, but now using the the white target light curve. For this, we use the transit model corresponding to the best fit transit parameters obtained using the new method for the respective white target light curve for each observation. 

We then use the common-mode trend and time both as GP regressors to fit the individual $\rm \lambda$LCs independently. The common-mode trend helps in modelling the largely wavelength independent high frequency systematics including the known odd-even effect described in Section \ref{sec:odd_even}. Using time as an additional GP regressor helps in modelling the smoother low frequency trend in the Target $\rm \lambda$LCs which is due to the changing airmass through the night and is wavelength dependent. Similar to the conventional method described in Section \ref{sec:ts_old}, for the new method as well we keep all the transit model parameters except the transit depth and the limb darkening coefficient for each $\rm \lambda$LC fixed to the best fit values derived from the new method for the corresponding white target light curve (tabulated in Table \ref{tab:bestfitparams}).    

It should be noted that both the new and conventional methods of fitting the spectroscopic light curve use the common-mode trend which means the resultant transmission spectra from both the methods are relative to the white light curve transit depth used to derive the common-mode. We also considered two further possible GP regressor combinations excluding the common-mode trend: 1) $\rm \lambda$LC and 2) time and $\rm \lambda$LC. We show the resultant transmission spectra overplotted with those from using common-mode and time as GP regressor in the Figure 1 in supplementary material. Figure 2 in supplementary material shows the difference in per bin BIC (derived from the GP likelihood) between the respective methods used to fit target $\rm \lambda$LC. Based on the average per bin $\Delta$BIC for all observations as seen in Figure 2 of supplementary material, we conclude that using common-mode and time are the best favoured GP regressor combination for fitting the target $\rm \lambda$LC. 

In principle, using comparison spectroscopic light curves as GP regressors would be preferable over using the common-mode as a GP regressor, but this would concretely depend on the comparison star itself. In the precise case of WASP-19b observations in this paper, the comparison star is 1.22 magnitude fainter as compared to WASP-19 which leads to worse fits (higher BIC as seen in Figure 2 of supplementary material). Hence, this forces us to go for the next best option, which is using the common-mode and time as GP regressors. We note that the transmission spectra for all the observations obtained from the new method using common-mode trend are consistent with those obtained using the best GP regressor combination excluding the common-mode trend: comparison $\rm \lambda$LC and time.  

We show the $\rm \lambda$LCs and their best fits obtained from both the conventional and the new method in Figures 3 to 10 in the supplementary material. The respective transmission spectra for each observation from both the methods are plotted in Figure \ref{fig:abs_tdepth_all} and tabulated in Table \ref{tab:r150_ts_targ} and \ref{tab:r150_ts_corr}. 

\subsubsection{Comparison between the transmission spectrum from the conventional and the new method}
\label{sec:compare_ts_method}

\begin{figure*}
  \centering
  \includegraphics[scale=0.45]{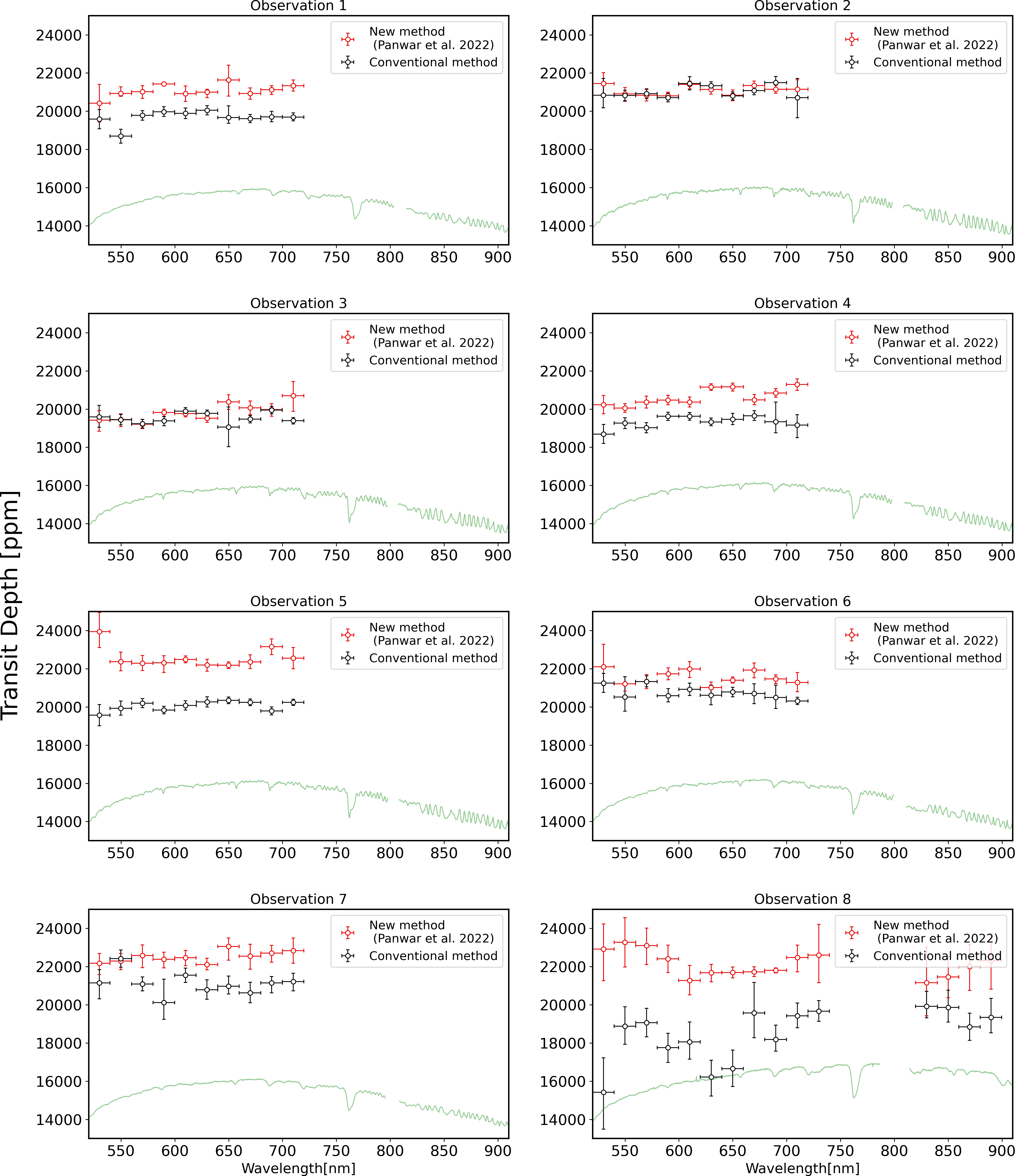}
  
  \caption[1]{Comparison of the transmission spectra of WASP-19b for each of the eight GMOS-R150 observations extracted using the conventional method (black points) and the new method (red points) from \citetalias{Panwar2022} described in Section \ref{sec:ts_old} and \ref{sec:ts_new} respectively. Observation numbers of each epoch are the same as in Table \ref{obsstats}. Observed GMOS-R150 stellar spectrum of WASP-19 for an arbitrary exposure is shown for each epoch in green. As described in Section \ref{sec:compare_ts_method} we eventually use the transmission spectra from the new method for subsequent interpretation in the paper.
  }
  \label{fig:abs_tdepth_all}
\end{figure*}

We compare the transmission spectrum at each epoch derived from two methods: the conventional method in which we use the comparison star $\rm \lambda$LCs followed by common-mode trend subtraction, and the new method in which we do not use the comparison star $\rm \lambda$LCs and use the common-mode trend as a GP regressor. The transmission spectrum for each epoch obtained from both the methods are plotted for comparison in Figure \ref{fig:abs_tdepth_all}. The average per bin precision on the transmission spectrum from the two methods are comparable. However, particularly in the case of observation 1 and observation 8 (the noisiest observations in our dataset), the new method yields 30 \% and 50 \% respectively smaller average transit depth uncertainties compared to the conventional method. The RMS of the residuals in $\rm \lambda$LC from the new method across all observations are smaller by a factor of 3 on average as compared to those from the conventional method (annotated in $\rm \lambda$LC Figures 1 to 8 in the supplementary material). The better transit depth precision yielded by the new method is an outcome of both not using the fainter and hence noisier comparison star $\rm \lambda$LCs and a generalized non-linear mapping of the white light curve common-mode trend with the individual $\rm \lambda$LCs. This advantage of the new method was also demonstrated in \citetalias{Panwar2022}. In essence, the transmission spectra from the new method are less susceptible to additional uncertainties and bias introduced in the conventional method by simply dividing the target spectroscopic light curves by spectroscopic light curves of a significantly fainter comparison star. Since we don't use the comparison star spectroscopic light curves in the new method, the transmission spectra hence obtained are not affected by wavelength dependent changes in the stellar spectra due to potential variability of the comparison star itself which could complicate our study and correction of the host star variability on the transmission spectrum of WASP-19b in Section \ref{sec:correct_stellar_activity}. Hence, in the subsequent sections in the paper, we consider only the transmission spectra from the new method for further interpretation. 

In Section \ref{sec:discussion} we also discuss the effect of stellar variability on the transmission spectrum during each epoch and introduce a new way to correct them before constructing the combined transmission spectrum and comparing it with previous studies and atmospheric models.


\section{Discussion} 
\label{sec:discussion}

\subsection{Effect of stellar activity on the transmission spectrum of WASP-19b}
\label{sec:stellar_activity_discuss}

WASP-19 is known to vary at a level of $\sim$ 2 \% peak to trough as seen from the TESS photometry and our ground based monitoring from LCO telescopes (Figure \ref{fig:LCO_phot}), which translates to $\sim$ 2 \% variation in white light curve transit depth from GMOS observations. We do not identify any spot crossing events in our GMOS observations like those observed by \cite{Espinoza2019} and \cite{Mancini2013} despite the precision of the GMOS transit light curves. However, spots and faculae are also expected to significantly affect the transmission spectrum via the transit light source effect (\citeauthor{Rackham2017}, \citeyear{Rackham2018}, \citeyear{Rackham2019}) especially in the visible wavelength range covered by our GMOS observations. Hence, it is necessary to correct for this effect of stellar activity in the transmission spectrum at each epoch first before combining them and producing the transmission spectrum.  

We estimate the impact on the transmission spectrum from unocculted stellar heterogeneity in a semi-empirical way. We use the estimates on temperature contrast and covering fraction of spots and faculae reported by \cite{Espinoza2019} based on the spot and faculae crossing events observed in their Magellan/IMACS light curves and previously by \cite{Mancini2013}. \cite{Espinoza2019} use the PHOENIX model stellar photospheres (\citealt{Husser2013}) and the observed spot and faculae contrasts to derive the estimates on spot and faculae covering fraction ranges that correspond to the 2 \% amplitude of stellar flux variability of WASP-19 seen in the visible bandpass. In Table \ref{tab:stellar_het_params} we summarize the spot and faculae properties from \cite{Espinoza2019} and \cite{Mancini2013} which we use in this paper to compute the effect of stellar activity on the transmission spectrum of WASP-19b.   

\begin{table*}
\caption[1]{Stellar photospheric heterogeneity parameters for WASP-19 corresponding to the peak to trough V band variability amplitude of 2 \%.}
\label{tab:stellar_het_params}
\centering

\begin{tabular}{llll}
\hline
Parameter & Description & Value & Reference \\
\hline

T$_{\mathrm{phot}}$ & Immaculate stellar photosphere temperature & 5460 K & \cite{Doyle2013} \\
T$_{\mathrm{spot,high}}$ & High contrast spot temperature & 4780 K & \cite{Mancini2013} \\ 
T$_{\mathrm{spot,low}}$ & Low contrast spot temperature & 5270 K & \cite{Espinoza2019} \\ 
T$_{\mathrm{fac}}$ & Faculae temperature & 5600 K & \cite{Espinoza2019} \\ 
\textit{f}$_{\mathrm{spot,high}}$ & High contrast spot covering fraction & 2$_{-0.7}^{+2.4}$ \% & \cite{Espinoza2019} \\ 
\textit{f}$_{\mathrm{spot,low}}$ & Low contrast spot covering fraction & 10$_{-5}^{+30}$ \% & \cite{Espinoza2019} \\
\textit{f}$_{\mathrm{fac}}$ & Faculae covering fraction & 19$_{-10}^{+31}$ \% & \cite{Espinoza2019} \\
\hline
\end{tabular}
\end{table*}


Now that we have an estimate of the properties of the stellar inhomogeneities on the stellar surface corresponding to the visible stellar flux variability, we estimate their impact on the theoretical transmission spectrum of WASP-19b. To do so, we use the open source atmospheric modelling code \texttt{platon} (\citealt{Zhang2019}, \citeyear{Zhang2020} ) based on \texttt{ExoTransmit} (\citealt{Kempton2016}) to calculate forward models for the transmission spectra corresponding to solar metallicity and C/O, accounting for the effect of unocculted stellar photospheric heterogeneity. The wavelength dependent effect of stellar variability implemented by \texttt{platon} (Equation 4 in \citealt{Zhang2019}) is the same as the one described in \cite{McCullough2014} and \cite{Rackham2018}. We use \texttt{platon} to calculate the forward models for three independent cases using the parameters listed in Table \ref{tab:stellar_het_params} : high contrast spots, low contrast spots, and faculae. Realistically, the stellar photosphere would be a combination of the three cases with spots and faculae contributing opposing effects. However, we consider the effect of each case separately to inspect the overall range of effect on the transmission spectrum due to stellar activity.      

\begin{figure*}
  \centering
  \includegraphics[scale=0.6]{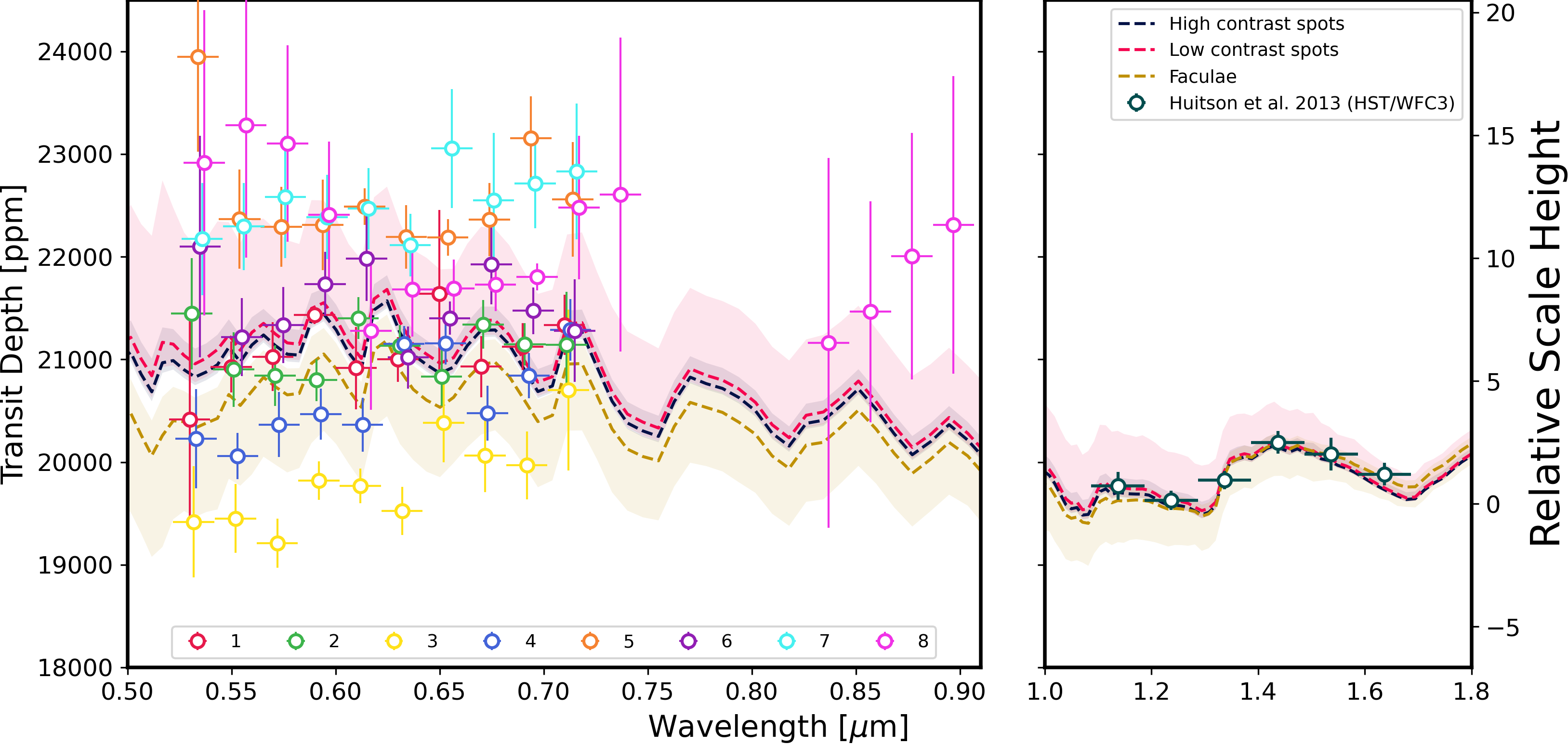}
  \caption[1]{Optical transmission spectra of WASP-19b from the individual 8 transit observations from Gemini/GMOS in this work in context of HST/WFC3 spectra from \cite{Huitson2013}. Overplotted for comparison are \texttt{platon} forward models matching the water absorption feature in HST/WFC3 for a cloud free atmosphere with solar metallicity and C/O and including the mean (dashed lines) and 1 $\sigma$ range (shaded) of the correction factor due to unocculted high contrast spots (in blue), low contrast spots (red) and faculae (yellow) as described in more detail in Section \ref{sec:stellar_activity_discuss}. All the \texttt{platon} forward models here have been normalized to the HST/WFC3 observations.    
  }
  \label{fig:TS_abs_hst}
\end{figure*}

In Figure \ref{fig:TS_abs_hst} we show the \texttt{platon} forward models normalized to the HST/WFC3 spectrum from \citealt{Huitson2013} and the individual GMOS transmission spectra from each of the eight observations overplotted. We notice from Figure \ref{fig:TS_abs_hst} that unocculted stellar spots and faculae corresponding to the contrasts and covering fraction ranges for WASP-19 as estimated by \citealt{Espinoza2019} can lead to an offset of up to $\sim$ 3000 ppm in the GMOS-R150 wavelength range of 520-900 nm. We measured this offset range from the same \texttt{platon} models overplotted in Figure \ref{fig:TS_abs_hst} which account for the effect due to unocculted high and low contrast spots and faculae with respect to the contrasts and covering fractions mentioned in Table \ref{tab:stellar_het_params}. We emphasise that this is consistent with the observed spread $\sim$ 4000 ppm seen in the mean levels of the transmission spectra from our GMOS observations (see Figure \ref{fig:TS_abs_hst}). 

\begin{figure}
  \centering
  \includegraphics[scale=0.42]{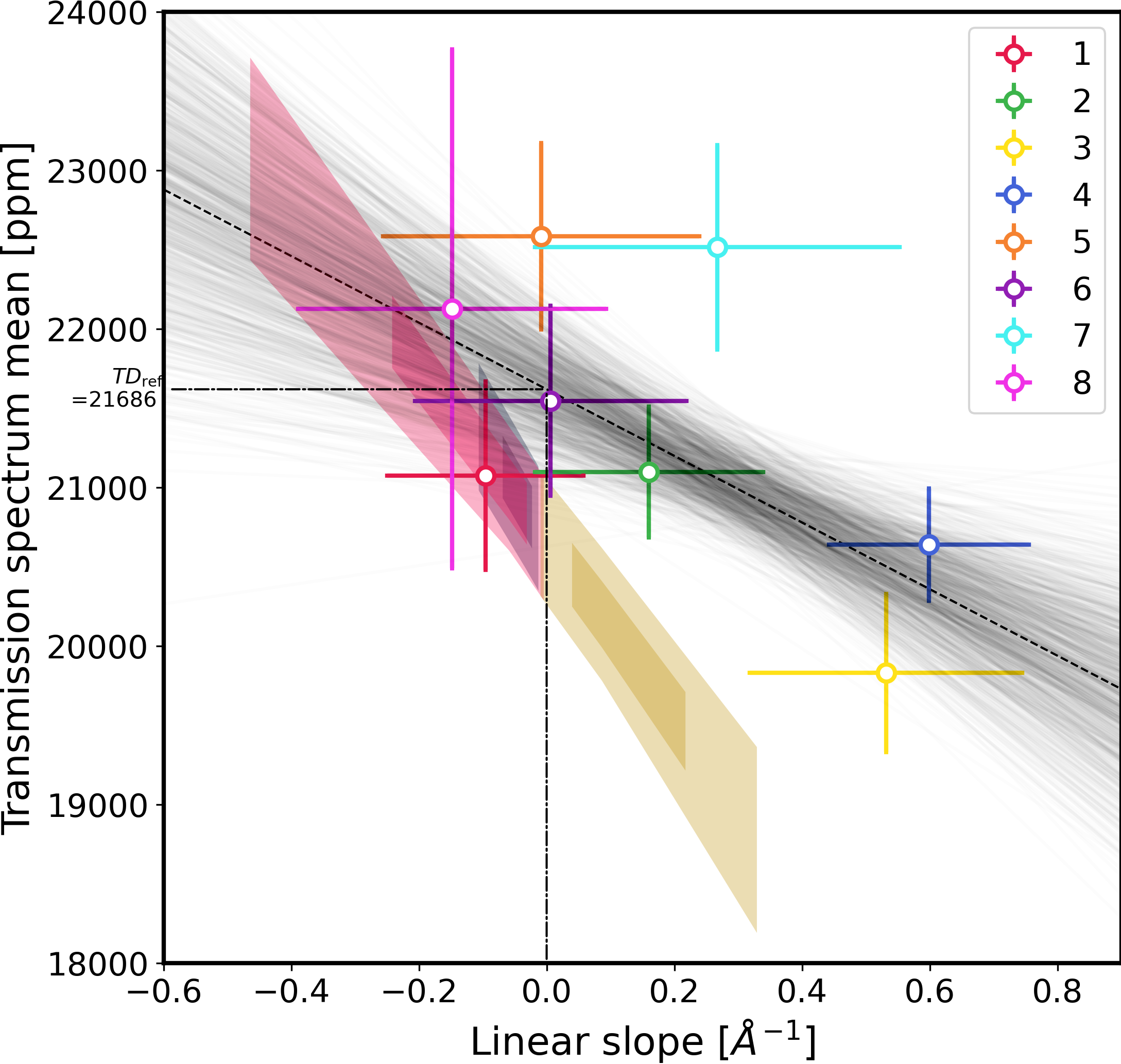}
  \caption[1]{Slope of the transmission spectra (X axis) obtained from 8 GMOS transit observations vs the mean level of each transmission spectra (Y axis). We observe an anticorrelation between the slope and the transmission spectral mean level with the corresponding Pearson correlation coefficient of --0.61 at 2-tailed $p$ value of 0.1. The linear fit to the slope and transmission spectra mean is shown as black dashed line along with grey lines showing fits from random samples from \texttt{emcee} posterior of the linear fit. The black dot-dash lines show the projection of the linear fit we use to obtain TD$_{\mathrm{ref}}$ in Section \ref{sec:correct_stellar_activity}. Overplotted in shaded regions are the $\pm$1$\sigma$ (dark shaded) and $\pm$2$\sigma$ (faint shaded) slope vs transmission spectrum mean derived from the \texttt{platon} models plotted in Figure \ref{fig:TS_abs_hst} which account for the effect of unocculted high contrast spots (blue), low contrast spots (red) and faculae (yellow).
  }
  \label{fig:TS_slope_vs_mean_level}
\end{figure}

\begin{figure*}
  \centering
  \includegraphics[scale=0.4]{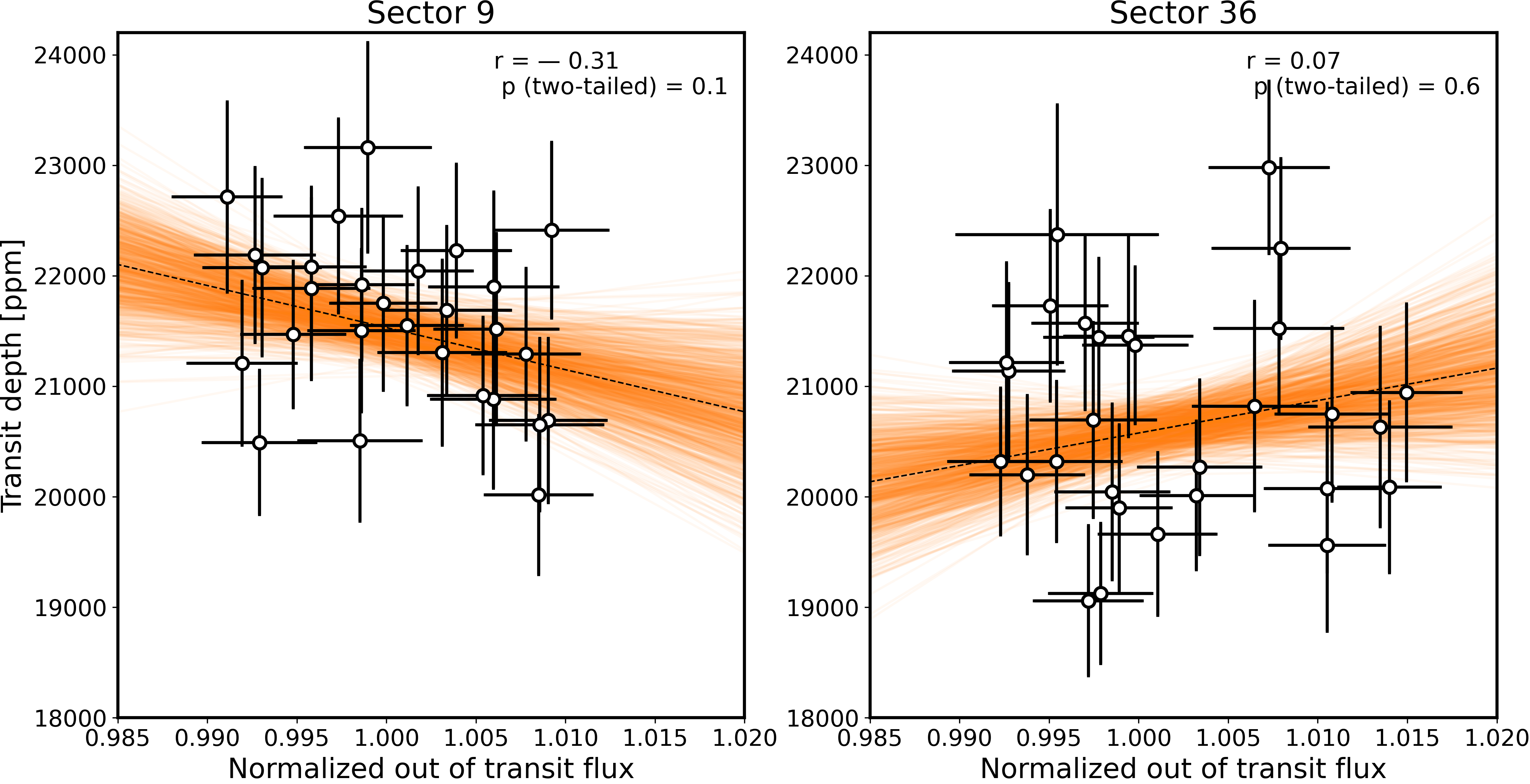}
  \caption[1]{Best fit TESS transit depths vs median out of transit flux for the 58 transits observed by TESS in Sector 9 (left panel) and Sector 36 (right panel). The transit depths were obtained by fitting the individual transit light curves as described in Section \ref{tess_lc_analysis}. The median out of transit light curve flux was measured after normalizing the two orbits of each sector by their respective median flux to mitigate any effect of the systematic offset between the orbits of individual sector. The dashed black line shows the linear fit to the points and the orange lines show randomly drawn samples from the MCMC posteriors of the linear fit. In the inset for each sector are the Pearson correlation coefficients and two tailed p values for the correlation between the transit depths and out of transit flux.}     
  \label{fig:tess_tdepth_oot}
\end{figure*}

We also inspect the TESS photometry for the variation of WASP-19b's transit depths across the 58 transits with respect to the stellar flux. We first compare the variation in absolute TESS SAP (simple aperture photometry) flux measured for WASP-19b and the two comparison stars and find that all three stars show a similar offset in their absolute SAP flux levels from Sector 9 to Sector 36. This implies that the change in SAP flux of WASP-19 between two sectors is not astrophysical. Hence, we conduct the transit depth vs out of transit flux comparison for the two sectors independently by normalizing each sector's photometry independently. Specifically we normalize each of the two orbits for both sectors by their respective median SAP flux, as shown in Figure \ref{tess_obs}. The resultant normalized out of transit flux vs transit depth comparison for both the sectors is shown in Figure \ref{fig:tess_tdepth_oot}. Within the individual sectors themselves WASP-19's flux varies by $\sim$2 \% which we interpret as due to rotational modulation by spots and faculae as also evident from the Lomb Scargle periodograms of both the sectors in Figure \ref{tess_obs}. Both the sectors show a scatter of $\sim$ 4000 ppm which is consistent with the spread in mean transmission spectra level seen in the eight GMOS-R150 observations (Figure \ref{fig:TS_slope_vs_mean_level}), and the 6 Magellan/IMACS transmission spectra from \cite{Espinoza2019}. This is expected because both the GMOS-R150, Magellan/IMACS, and TESS observations have a significant overlap in wavelength range.

We find that Sector 9 photometry shows an anticorrelation between the out of transit stellar flux and the transit depth as expected from unocculted spots. The Pearson correlation coefficient of --0.31 at two tailed p value of 0.1 indicates that the anticorrelation is not significant. We speculate that spot or faculae occultations by the planet during the transits observed by TESS could be responsible for this deviation from the anticorrelation expected from the stellar brightness variations due to only unocculted spots or faculae. Both spot and faculae occultations have been observed by \cite{Espinoza2019} with photometric amplitudes $\sim$ 3000 ppm in the transit light curves. The average RMS we obtain from the TESS light curves is of the same order of $\sim$3000 ppm as shown in Figures 13 and 14 in the supplementary material. Hence, from our fits of each TESS transit light curve in this work it is not possible to detect and fit for the signatures of spot or faculae occultations along with the transit signal. Hence, we speculate that the TESS transit depths are overall affected by spot and faculae occultations.

Unocculted spots and faculae impart not only an offset but also a slope to the optical transmission spectrum which can vary significantly across multiple epochs due to stellar variability. We demonstrate this in the forward models plotted in Figure \ref{fig:TS_abs_hst}. On average, high and low contrast spots impart a positive offset and a negative slope, while faculae, on the other hand impart a negative offset and a positive slope on the spectrum (\citealt{Rackham2017}, \citealt{Espinoza2019}). To measure this effect at first order on the GMOS transmission spectrum at each epoch, we fit a linear slope to the transmission spectrum from each observation and compare the best fit linear slope value to the corresponding mean transmission spectrum level for each epoch. The linear slopes and mean transmission spectrum level for each epoch are plotted in Figure \ref{fig:TS_slope_vs_mean_level}. A visual illustration of how we construct Figure \ref{fig:TS_slope_vs_mean_level} is given in Figure \ref{fig:slope_off_illustrate}. We find that the GMOS observations with larger mean transmission spectrum level have a negative slope and vice-versa. We measure an anticorrelation between the transmission spectra mean level and their slopes with Pearson correlation coefficient --0.61 at 2-tailed $p=0.1$. This anticorrelation is expected from the theoretical forward models accounting for the effect of stellar variability as we demonstrate next. 

For comparison with predictions from theoretical models, we compute the \texttt{platon} forwards models for the transmission spectrum of WASP-19b while also accounting for the effect of unocculted spots and faculae. These are the models plotted in Figure \ref{fig:TS_abs_hst}. To obtain the expected slopes and offsets from the \texttt{platon} forward models we follow the same approach as that applied to the observed transmission spectra. In the GMOS wavelength range of 520 to 720 nm (common to all 8 epochs) we fit a linear slope to the \texttt{platon} forward models corresponding to mean and $\pm$1$\sigma$ and $\pm$2$\sigma$ properties of the three cases from Table \ref{tab:stellar_het_params}: high-contrast spots, low-contrast spots and faculae. The predicted mean and $\pm$1$\sigma$ and $\pm$2$\sigma$ slope and transmission spectrum mean level for all three cases are shown as shaded regions in Figure \ref{fig:TS_slope_vs_mean_level}. 

We find that the transmission spectra mean level vs slope trend in GMOS observations is broadly consistent with the predictions from the forward models that account for the stellar variability due to spots and faculae, as shown by the shaded region in Figure \ref{fig:TS_slope_vs_mean_level}. We note that the model predicted trends deviate from the best fit linear trend to the data, especially at the faculae end, as shown in Figure \ref{fig:TS_slope_vs_mean_level}. This is because the models we consider describe end-member effect from a purely spot or faculae dominated stellar photosphere. Realistically, WASP-19's stellar photosphere is more likely to host a mix of both spots and faculae. This explains the deviation between the slope vs offset trend predicted by the models as shown by the shaded region in Figure \ref{fig:TS_slope_vs_mean_level} and the trend measured in the data. Nevertheless, the trend in slope vs offset space across all epochs can have implications on the morphology of the final transmission spectra combined from multiple epochs. Corrections for both the slope and offset at each epoch need to be applied before combining the transmission spectra. We use the observed trend in transmission spectral slopes and offsets to combine multi-epoch spectra; we describe this new empirical approach to correct for the effect of stellar variability in the following section.

\subsection{A new empirical approach to correct for stellar variability across multiple epochs}
\label{sec:correct_stellar_activity}
Conventionally, transmission spectra obtained at different epochs have been combined by first applying an offset with respect to a reference level, e.g., as done for WASP-19b by \cite{Espinoza2019}. This offset is constant with respect to wavelength. For comparison, we also first apply this constant offset correction to our GMOS observations. We choose a reference transit depth TD$_{\mathrm{ref}}$ = 21686 ppm which is the value on the Y axis projected by the linear fit to the transmission spectral mean and slope corresponding to zero transmission spectrum slope on the X axis. This projection is marked as black dot dashed lines in Figure \ref{fig:TS_slope_vs_mean_level}. Note that the TD$_{\mathrm{ref}}$ measured from the linear fit to the measured slopes and offsets of the data do not coincide with the analogous `zero-point' indicated by the predicted slopes and offsets from the \texttt{platon} models marked by the shaded region in Figure \ref{fig:TS_slope_vs_mean_level}. This is in part related to the normalization of the \texttt{platon} models to the HST/WFC3 data. Another reason for this deviation is, as we mention in Section \ref{sec:stellar_activity_discuss}, the \texttt{platon} models we use represent only spot or only faculae dominated photosphere. Either a mixture of both spot and faculae on the photosphere or a change in normalization of the models, or both can change the nature of the linear trend between the transmission spectral slopes and offsets.       

\begin{figure*}
  \centering
  \includegraphics[scale=0.5]{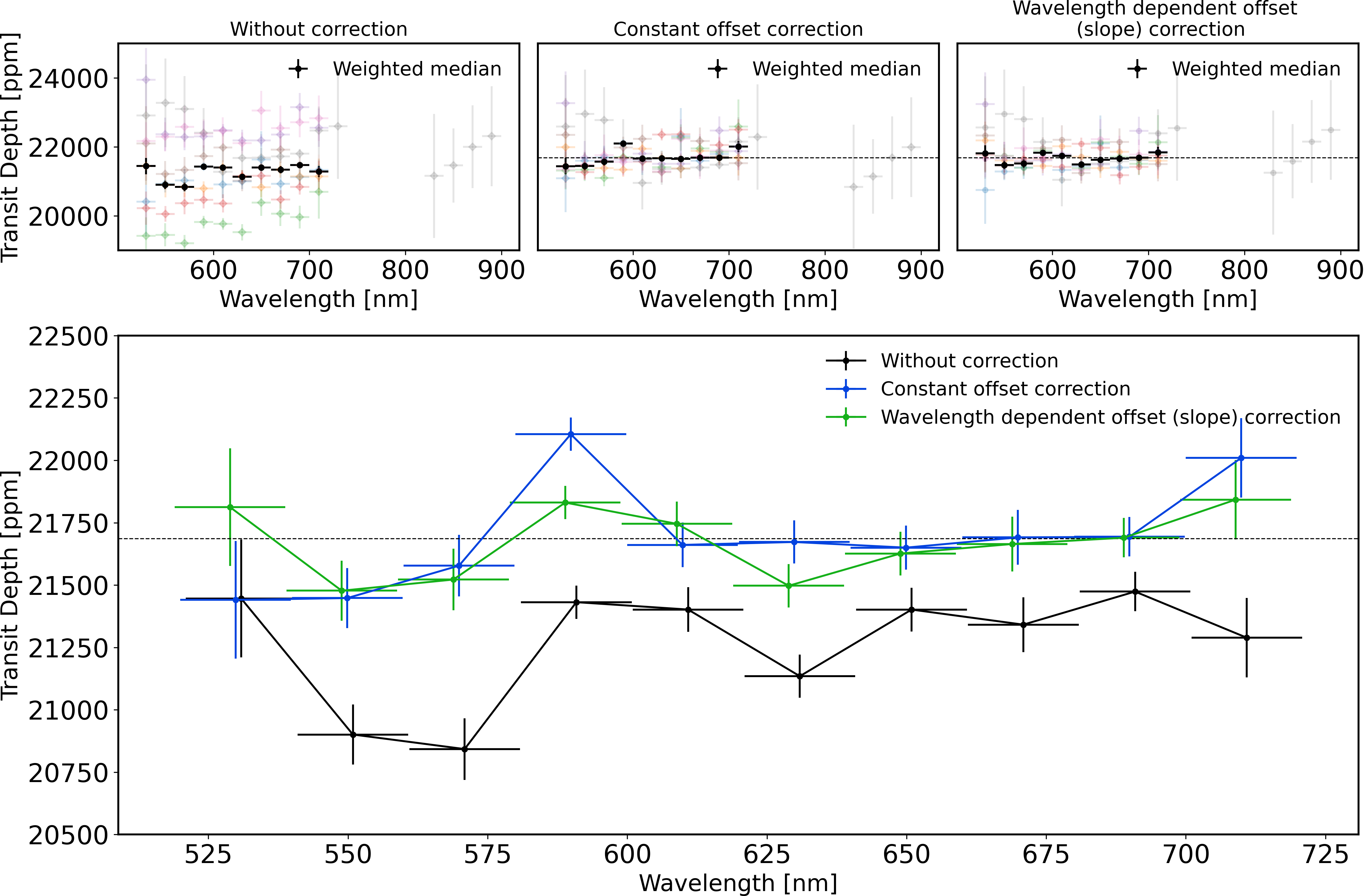}
  
  \caption[1]{Comparison of transmission spectrum of WASP-19b obtained from weighted median combining the transmission spectrum from 8 epochs in three different ways described in more detail in Section \ref{sec:correct_stellar_activity} and shown here in top three panels: 1) without applying any slope or offset corrections to the individual epochs, 2) applying a constant offset correction to the individual epochs, and 3) applying a wavelength dependent offset (slope) correction to the individual epochs. Coloured points in all top three panels show the transmission spectrum measured at each epoch and the weighted median combined spectrum in black points. The combined transmission spectrum from 1), 2), and 3) are overplotted for comparison in the bottom panel in black, blue, and green points respectively. Note that not applying any stellar variability corrections leads to spurious features in the transmission spectrum as seen in the black points, which are corrected to some degree by constant offset correction as seen in the blue points, and to a better degree with wavelength dependent offset correction as seen in the green points.
  }
  \label{fig:GMOS_stellar_contamination_correction}
\end{figure*}

Following an empirical approach we choose to use the TD$_{\mathrm{ref}}$ measured from the data. We first apply a constant offset to the GMOS transmission spectrum from each epoch with respect to TD$_{\mathrm{ref}}$ before weighted median combining them to obtain the combined transmission spectrum. The combined transmission spectrum after constant offset correction is shown as black points in Figure \ref{fig:GMOS_stellar_contamination_correction}. 

However, it is clear from the anticorrelation observed between the transmission spectra means and slopes in Figure \ref{fig:TS_slope_vs_mean_level} that just a constant offset correction is not enough as it does not remove the different slopes imparted by stellar variability at each epoch. Hence, instead of a constant offset correction, we introduce here a new empirical approach that also corrects for the slope. For a transmission spectrum we compute the difference between its linear fit and the TD$_{\mathrm{ref}}$ for each wavelength bin. The wavelength dependent offset hence obtained for each epoch can now be used to correct both the slope and offset in a transmission spectrum. We apply the wavelength-dependent offset to the transmission spectra at each epoch and then weighted median combine the slope corrected spectra to obtain a combined transmission spectrum, shown in Figure \ref{fig:GMOS_stellar_contamination_correction} as green points.  

We emphasize that there are some major caveats to this empirical approach of slope and offset correction. Our approach is agnostic to the spectral slope present in the spectra due to the planetary atmosphere itself. If a spectral slope intrinsically due to the planetary atmosphere exists (e.g. due to haze scattering), it would vanish after our slope correction. Hence, our approach cannot resolve the discrepancy between the Magellan/IMACS and VLT/FORS2 data with respect to presence or absence of hazes. Moreover, given the wavelength coverage of the GMOS-R150 data, the GMOS transmission spectrum in this work is less sensitive to a slope due to hazes which impart a much stronger signature blueward of 400 nm. However, what our approach of slope and offset correction preserves is any prominent spectral features in the individual transmission spectra. In other words, our slope correction would remove any planetary atmospheric slope, but will retain spectral features, e.g. due to Na and K or TiO/VO molecular bandheads if present. Especially in the wavelength range of 520 to 720 nm probed by our GMOS-R150 observations, we expect a stronger contribution from spectral features from Na/K or TiO/VO as compared to scattering due to hazes. We next compare the stellar variability corrected (for both slope and offset) combined GMOS transmission spectrum with atmospheric forward models. Another caveat of our approach is that the linear approximation of the impact of unocculted spots in terms of an offset and slope works well given the precision and wavelength span of our data. \cite{McCullough2014} show that the impact of spots when modelling the spot and quiescent photosphere spectra as blackbodies is in general non-linear with respect to wavelength. Hence, we recommend exploring other parametric approximations e.g. a quadratic polynomial, which might be better suited for modelling the effect of unocculted stellar spots in the transmission spectra obtained from datasets spanning different wavelength ranges and precisions. In summary, we recommend correcting the impact of heterogeneous stellar photosphere in the transmission spectrum at each individual epoch before combining them.

\subsection{Optical and near-infrared transmission spectrum of WASP-19 and comparison with forward atmospheric models}
\label{sec:atmosphere_interpretation}

We now discuss the combined GMOS transmission spectrum which has been obtained after correcting for stellar variability in conjunction with the HST/WFC3 spectrum and their comparison to the forward models for WASP-19b's atmosphere computed using \texttt{platon}. We restrict our comparison to the 520 to 720 nm range for the GMOS transmission spectrum as the only data points we have beyond 720 nm are from observation 8 with much larger uncertainties for any meaningful model comparison. We expect that given the limited wavelength range and resolution of the transmission spectrum per epoch, an atmospheric retrieval wouldn’t be able to meaningfully resolve the degeneracies between the contribution due to stellar variability which causes the offset between the GMOS and HST/WFC3 data, and that from the planetary atmosphere. Hence, we choose to perform only forward model comparisons which we expect are sufficient for our goal of testing the presence or absence of TiO and Na features. 

We normalize the HST/WFC3 transmission spectrum from \cite{Huitson2013} to the TD$_{\mathrm{ref}}$ calculated in Section \ref{sec:correct_stellar_activity}. We construct forward models using \texttt{platon} for five different cases, each with equilibrium chemistry and solar C/O: 1) Solar metallicity, 2) Solar metallicity and TiO abundance suppressed 100$\times$, 3) Solar metallicity and TiO abundance suppressed 1000$\times$ 4) Solar metallicity and no TiO, and 5) Solar metallicity, no TiO, and no Na. We apply a similar treatment of slope and offset removal to all the models in the GMOS bandpass as done for the GMOS transmission spectrum in Section \ref{sec:correct_stellar_activity}. For each \texttt{platon} model, we perform a linear fit to the model in the 520 to 720 nm range and calculate wavelength-dependent offsets with respect to the median of the model. We then apply these wavelength dependent offsets to the models additively in exactly the same manner as done for the GMOS spectra. We subsequently use these slope corrected transmission models for comparison with the observed transmission spectrum.  

Since the HST/WFC3 spectrum was obtained at a different epoch relative to the GMOS data, the stellar activity level is likely different between these epochs. Therefore, the relative offset between the GMOS and HST/WFC3 spectrum is arbitrary and needs to be accounted for when comparing the GMOS and HST/WFC3 spectrum together with the forward models. We leverage the shape of the spectral features in HST/WFC3 transmission spectra for comparison with the models. Each of the models we consider are consistent with the shape of the HST/WFC3 water absorption spectral feature from \cite{Huitson2013}. Hence, we first anchor all the \texttt{platon} models to match the HST/WFC3 points. Next, to compare the GMOS spectrum with the models, we apply a range of constant offsets (with respect to wavelength) in steps of 10 ppm to compute the minimum reduced chi-squared ($\chi^2_\nu$) between the GMOS spectrum and each model. Considering 11 degrees of freedoms (10 data points and 1 vertical direction offset), we find the minimum $\chi^2_\nu$ values for the five forward models as shown in Table \ref{tab:model_comparison} which we further use for model comparison.

\begin{table*}
\caption[1]{Model comparison criteria for arbitrarily offset GMOS data and the \texttt{platon} forward models shown for various cases shown here in the column `Model' and plotted in Figure \ref{fig:GMOS_hst_wfc3_platon_models} The top and bottom part of the table are for transmission spectrum with 20 nm and 10 nm wide bins respectively. The columns `$\chi^2_\nu$' and `BIC' show the minimum reduced chi-squared and the corresponding Bayesian Information Criterion respectively obtained by offsetting the GMOS transmission spectrum by varying amounts with respect to the \texttt{platon} forward model normalized to HST/WFC3. `N $\sigma$...' shows the number of sigmas by which each model is preferred over the Solar metallicity case. All the models have the metallicity and C/O ratio fixed to the solar value.}

\label{tab:model_comparison}
\centering

\begin{tabular}{llll}
\hline
Model & $\chi^2_\nu$ & N $\sigma$ from Solar & BIC \\
\hline
Solar metallicity & 4.298 & -- & 36.408\\
Solar metallicity and TiO abundance suppressed 100$\times$ & 1.794 & 4 & 21.189\\
Solar metallicity and TiO abundance suppressed 1000$\times$ & 0.954 & 4.5 & 18.264\\
Solar metallicity and no TiO & 0.876 & 5 & 17.869 \\
Solar metallicity, no TiO, and no Na & 1.654 & 3 & 23.995\\
\hline
\hline
Model & $\chi^2_\nu$ & N $\sigma$ from Solar and no TiO & BIC \\
\hline
(10 nm) Solar metallicity and no TiO & 2.771 & -- & 49.873\\
(10 nm) Solar metallicity, no TiO, and no Na & 3.380 & 3 & 63.800\\
\hline
\end{tabular}
\end{table*}

Based on the $\chi^2_\nu$, we rule out solar metallicity atmosphere with solar TiO abundance as compared to no TiO case at 5$\sigma$. A solar metallicity atmosphere with 1000$\times$ or completely depleted TiO best explains the shape of the GMOS transmission spectrum. This is 10 times lower TiO abundance reported from the FORS2 observations by \cite{Sedaghati2021}. The TiO depletion could be because of cold-trapping processes condensing TiO at the terminator as discussed by \cite{Parmentier2013} which could also explain the non-detection of Fe by \cite{Sedaghati2021}. As compared to models with no Na, we favour the models with solar abundance Na by 3$\sigma$ when considering the transmission spectrum for smaller 10 nm wide bins near the Na feature. A zoom of the models around the Na 589 nm doublet showing this tentative detection of Na is shown in Figure \ref{fig:GMOS_hst_wfc3_platon_models_zoomNa}. Interestingly, \cite{Sedaghati2017} also obtain a 3.4$\sigma$ Na detection in the VLT/FORS2 spectra, however the amplitude of the tentative Na absorption we detect in the GMOS spectrum is smaller than the VLT/FORS2 spectrum by $\sim$500 ppm.
\begin{figure*}
  \centering
  \includegraphics[scale=0.6]{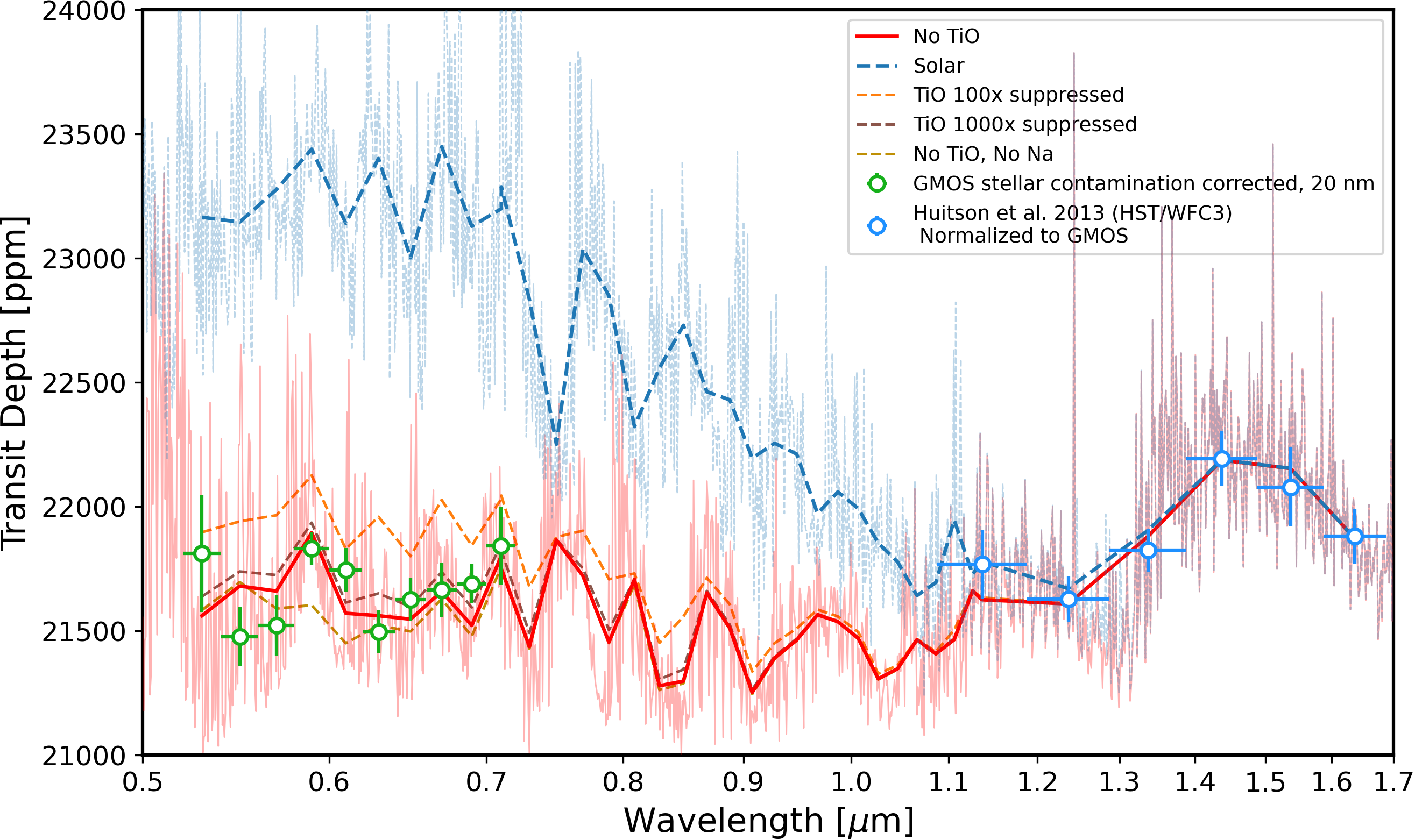}
  \caption[1]{The GMOS combined transmission spectrum of WASP-19b corrected for stellar variability and the HST/WFC3 transmission spectrum from \cite{Huitson2013} normalized to the median of the GMOS spectrum as compared to various \texttt{platon} forward transmission models binned in the GMOS and HST/WFC3 wavelength bins. We also show the higher resolution \texttt{platon} forward models for the `No TiO' and `Solar' scenarios. Similar to the stellar variability correction applied to the GMOS transmission spectrum, wavelength dependent offsets with respect to a linear fit only in the GMOS bandpass (520 to 720 nm) have been applied to the \texttt{platon} transmission models as well for a fair comparison. The HST/WFC3 transmission spectrum amplitude is consistent with solar H$_{2}$O abundance which is the same for all models plotted here. The shape of the GMOS spectrum is inconsistent with the shape predicted by the solar abundance models (dashed blue line) at more than 3$\sigma$ as compared to the models with suppressed TiO abundance, as discussed in more detail in Section \ref{sec:atmosphere_interpretation}.      
  }
  \label{fig:GMOS_hst_wfc3_platon_models}
\end{figure*}

\begin{figure}
  \centering
  \includegraphics[scale=0.4]{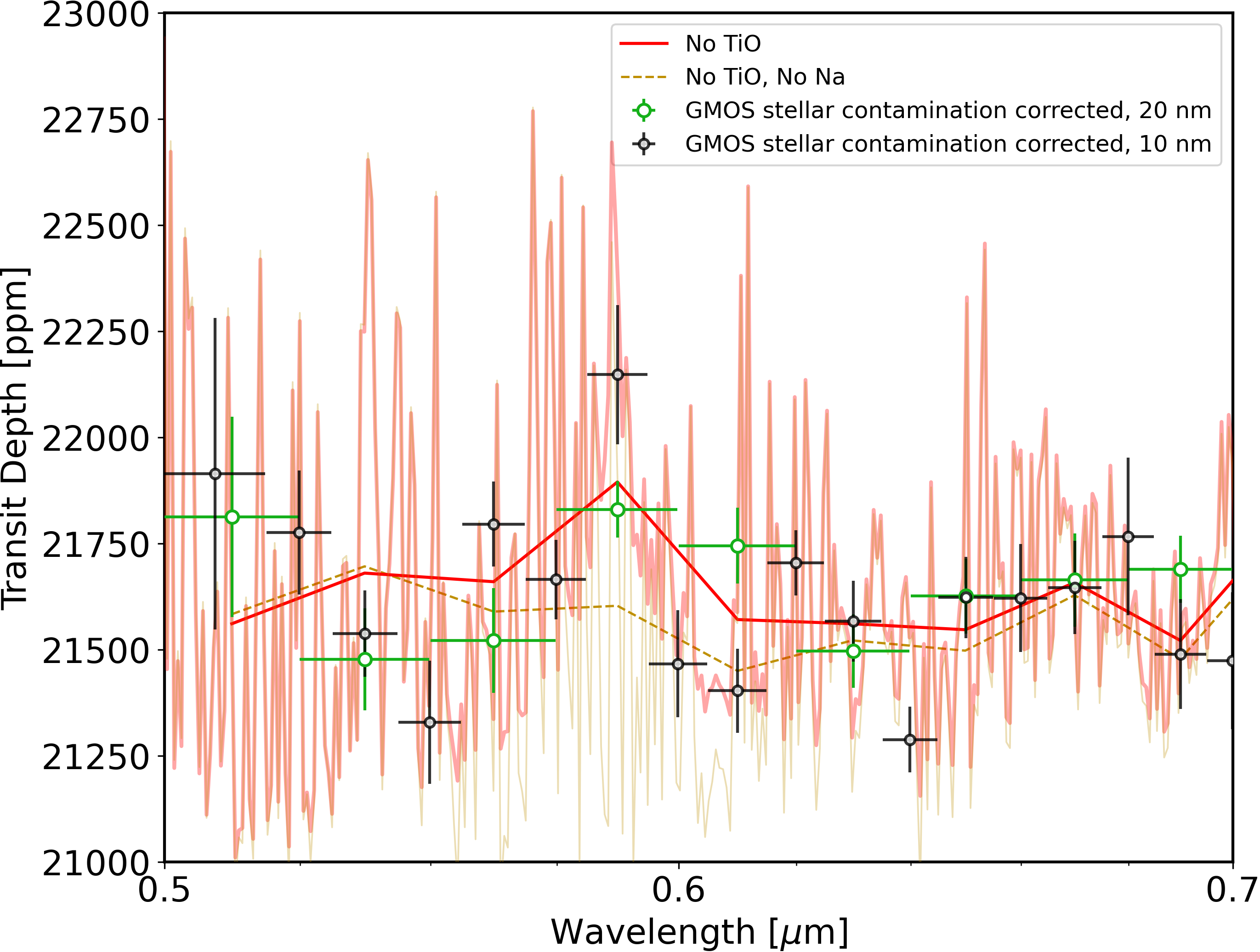}
  \caption[1]{A zoom-in of the GMOS combined transmission spectrum around the Na doublet at 0.589 $\mu$m. The green points and black points show the transmission spectrum computed for 20 nm and 10 nm wide bins respectively. We find that the 10 nm bin GMOS transmission spectrum favours a model with Na at $\sim$3$\sigma$ as compared to models with no Na indicating a tentative detection of Na absorption. We show the \texttt{platon} forward models used for comparison here in red for the model with Na and in yellow for the model with Na (thicker line showing the binned spectrum and thinner line showing the higher resolution model). 
  }
  \label{fig:GMOS_hst_wfc3_platon_models_zoomNa}
\end{figure}


\subsection{Comparison of the GMOS transmission spectrum of WASP-19b with previous studies}

\label{sec:comparison_prev_studies}

In this section, we compare our GMOS transmission spectrum of WASP-19b with that from Magellan/IMACS and VLT/FORS2. We first briefly summarize the results on the optical transmission spectrum of WASP-19b from two previous studies. 
The VLT/FORS2 transmission spectrum of WASP-19b from \cite{Sedaghati2017} was obtained by combining 3 transits observed at 3 different epochs. They detect significant features due to absorption by Na, H$_{2}$O, and substantially sub-solar TiO abundance, and a steep slope due to haze scattering towards the blue optical end of the spectrum. With further reanalysis of the same dataset using retrieval models accounting for the effect of stellar activity \cite{Sedaghati2021} arrive at a similar detection of 100$\times$ sub-solar TiO albeit at a lower significance (7.7$\sigma$ $\rightarrow$ 4.7 $\sigma$) of detection as compared to \cite{Sedaghati2017}. From their ESPRESSO high resolution spectroscopy observations \cite{Sedaghati2021} find a tentative 3 $\sigma$ detection of TiO consistent with 1000$\times$ sub-solar TiO abundance. However, as also noted by \cite{Sedaghati2021}, this isn't a confirmation of the VLT/FORS2 result due to low statistical significance.   

In a separate study of WASP-19b, \cite{Espinoza2019} observe 6 transits in the optical using Magellan/IMACS and do not find any signatures of absorption due to TiO, Na, or a scattering slope towards blue optical. \cite{Espinoza2019} observe spots and faculae crossings in two of their observations, along with a slope in the transmission spectrum from one of their transits attributed to the transit light source effect due to unocculted stellar spots. \cite{Espinoza2019} exclude the spectrum with the steep slope and construct the combined transmission spectrum from the other five epochs. Using a semi-analytical retrieval, \cite{Espinoza2019} conclude that the combined Magellan/IMACS transmission along with HST/WFC3 and Spitzer measurements are best explained by an atmosphere with solar composition water and sub-solar TiO and Na.     

We suggest that tension between the Magellan/IMACS spetrum (\cite{Espinoza2019}) and the VLT/FORS2 spectrum (\citeauthor{Sedaghati2017}, (\citeyear{Sedaghati2017}, \citeyear{Sedaghati2021})), can be explained by the stellar variability of the host star. \cite{Espinoza2019} and \citeauthor{Sedaghati2017}, (\citeyear{Sedaghati2017}, \citeyear{Sedaghati2021}) study transit observations taken at different epochs with likely varying levels of stellar variability. Note that given the wider wavelength range coverage of both Magellan/IMACS and VLT/FORS2 observations as compared to our GMOS observations, both studies account for the stellar variability contribution at individual epochs using a Bayesian retrieval that fits for both planetary atmospheric parameters and stellar photospheric heterogeneity parameters. \cite{Espinoza2019} find a statistically stronger contribution from stellar activity compared to a flat line in only one out of their six epochs which they eventually omit when constructing the final combined transmission spectrum using a constant offset correction. \cite{Sedaghati2021} from their \texttt{POSEIDON} retrieval of the three VLT/FORS2 epochs independently as well as jointly confirm the significant contribution from the stellar variability.

\begin{figure*}
  \centering
  \includegraphics[scale=0.6]{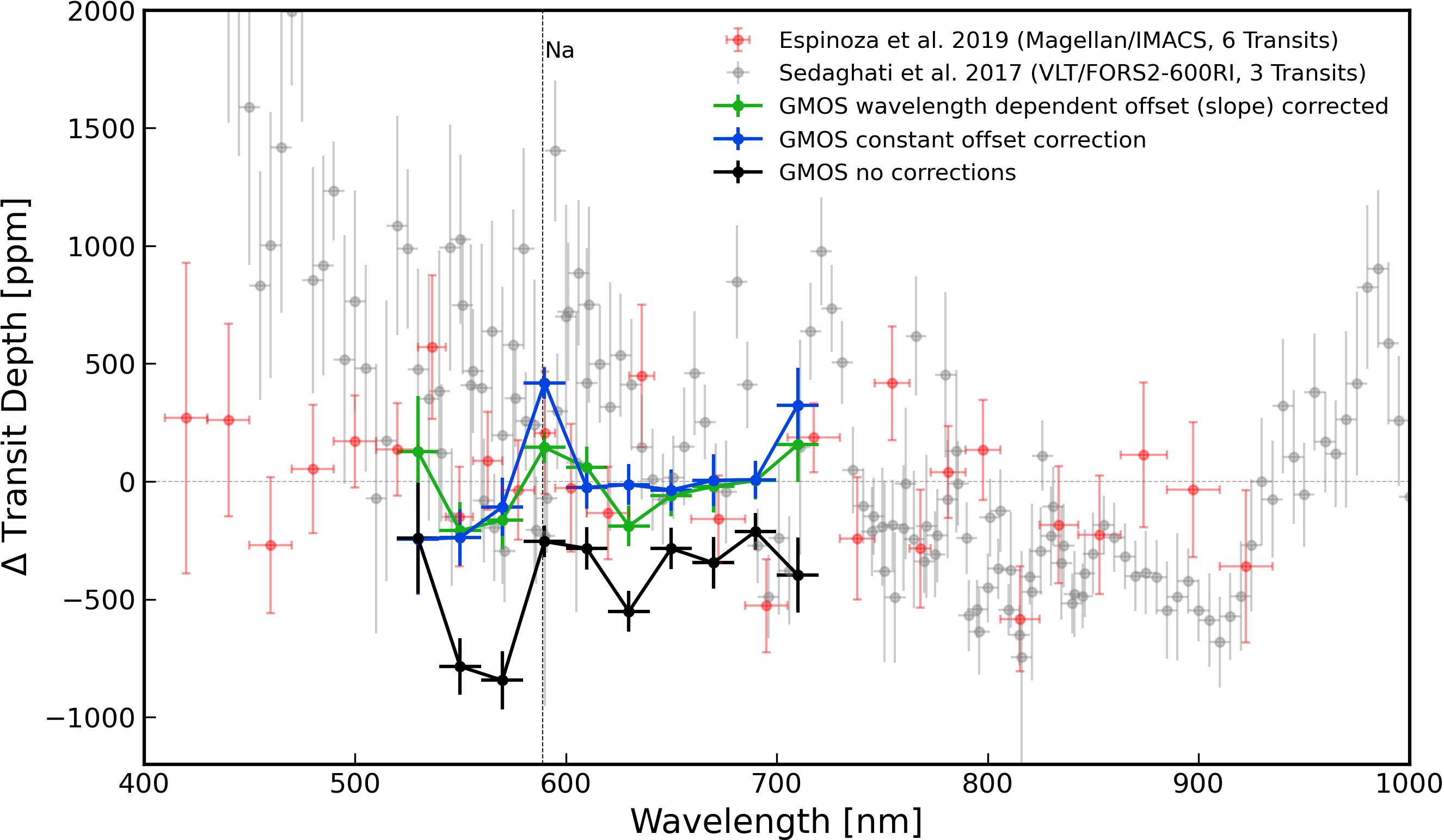}
  
  \caption[1]{Comparison of the mean subtracted combined GMOS transmission spectrum with and without any stellar variability corrections (black, blue, and green points) of WASP-19b with the mean subtracted combined transmission spectrum from Magellan/IMACS reported by \cite{Espinoza2019} (red points) and VLT/FORS2 reported by \cite{Sedaghati2017} (grey points). The stellar variability corrected GMOS spectrum (green and blue points) does not show any significant TiO absorption features in the 550 to 720 nm which is consistent with the Magellan/IMACS spectrum but inconsistent with the the VLT/FORS2 spectrum. The Gemini/GMOS spectra shows tentative evidence of Na absorption at lower amplitude as compared to \cite{Sedaghati2017}. However, the Gemini/GMOS, Magellan/IMACS, and VLT/FORS2 spectra all rule out the presence of solar TiO and are consistent with sub-solar TiO scenario at the terminator of WASP-19b.     
  }
  \label{fig:compare_sedaghati_espinoza}
\end{figure*}

Our GMOS observations exceed in both the telescope collecting area and the total number of epochs of the planet probed as compared to \cite{Espinoza2019} and \citeauthor{Sedaghati2017} (\citeyear{Sedaghati2017}, \citeyear{Sedaghati2021}). We obtain a median precision of 100 ppm per 20 nm bins, compared to 250 ppm per 10 nm bins for both Magellan/IMACS and VLT/FORS2 transmission spectra. Despite the smaller wavelength coverage, our GMOS observations spread over eight epochs probe a larger range of stellar variability, and at higher precision on the transmission spectrum both at individual epochs and the combined transmission spectrum. We leverage these aspects in the application of our empirical approach of applying relative corrections to individual epochs as described in Section \ref{sec:correct_stellar_activity}. We compare our stellar variability corrected spectrum with the Magellan/IMACS and VLT/FORS2 spectrum in Figure \ref{fig:compare_sedaghati_espinoza}. We observe that the GMOS spectrum without stellar correction (black points in Figure \ref{fig:compare_sedaghati_espinoza}) shows a feature in the 550 to 650 nm range at similar amplitudes to that seen in \cite{Sedaghati2017}. However, the feature diminishes significantly when we correct for stellar variability as seen in the green points in Figure \ref{fig:compare_sedaghati_espinoza}. 

From our stellar variability corrected spectrum, we conclude that we do not observe significant absorption features due to TiO, which is consistent with the findings of \cite{Espinoza2019} from their Magellan/IMACS spectrum. However, our tentative detection of Na absorption is inconsistent with \cite{Espinoza2019} and more consistent with \cite{Sedaghati2017}. Given our weak evidence for Na absorption, this doesn't confidently resolve the tension between \cite{Espinoza2019} and \cite{Sedaghati2017} with respect to Na. Additionally, our GMOS spectrum favours 1000$\times$ or lower sub-solar TiO scenarios and rules out the solar TiO scenario (see Section \ref{sec:atmosphere_interpretation}) at high level of significance. This is consistent with the findings of both \cite{Espinoza2019} and \citeauthor{Sedaghati2017}, (\citeyear{Sedaghati2017}, \citeyear{Sedaghati2021}) who also rule out the solar TiO scenario from their Magellan/IMACS and VLT/FORS2 spectrum respectively. Interestingly, similar to our GMOS data, the ESPRESSO high resolution observation of \cite{Sedaghati2021} is also consistent with 1000$\times$ sub-solar TiO. We emphasize that despite the differences in spectral morphology, the depletion of TiO at the terminator of WASP-19b is a scenario that explains the transmission spectrum from the GMOS, Magellan/IMACS, VLT/FORS2, and ESPRESSO observations.

\section{Conclusions}
\label{conclusions}

We present the optical transmission spectrum for the ultra-hot Jupiter WASP-19b from eight transits of the planet observed using GMOS on Gemini South spread over a span of two years. The main conclusions from our study are:

1) To extract the transmission spectrum of WASP-19b, we mitigate the effects of a fainter comparison star using the method to analyse MOS data that is presented in \cite{Panwar2022}. Using this method, we obtain an average factor of three improvement in the RMS of the spectroscopic light curves compared to the conventional method. We measure the transmission spectra without introducing additional uncertainties from a faint comparison star. We use a Bayesian framework for propagating uncertainties when modelling the systematics in the target star light curves. 

2) We find that the transmission spectra of WASP-19b obtained at different epochs vary suggestively in terms of their slopes and relative offsets which impedes the process of combining them to construct the final transmission spectrum. We interpret this as a result of the impact of stellar variability on the transmission spectra at individual epochs. WASP-19b orbits an active solar type star which shows a stellar flux variability of 2\% in the optical as confirmed by both TESS and ground-based broadband photometry. Hence, relative corrections for stellar variability at each epoch need to be applied before constructing the combined transmission spectrum.

3) We observe that the effect of stellar variability manifests broadly in two ways: a slope and an offset to the transmission spectra, both of which need to be measured and accounted for when co-adding multiple epochs. We compute these effects for WASP-19b's transmission spectrum using the spot (positive slope) and faculae (negative slope) temperature contrasts and covering fractions corresponding to the host star's amplitude of variability as measured by previous studies.

4) For the eight GMOS observations of WASP-19b presented in this paper, the offsets between the transmission spectra from each epoch span a range of $\sim$4000 ppm, and a slope -0.4 to 0.6 ppm per Angstrom. The trend in the transmission spectra slope vs offset broadly matches in amplitude and sign that predicted by forward atmospheric models accounting for the effect of stellar spots and faculae.

5) We introduce a new empirical approach for correcting the stellar variability in transmission spectra across multiple epochs by using the measured spectral slopes and offsets to apply relative corrections between epochs and to construct the combined transmission spectrum.    

6) Our stellar variability corrected GMOS spectrum rules out the solar TiO scenario at 5$\sigma$, and is most consistent with the 1000$\times$ sub-solar TiO or lower TiO abundance scenario. Significant depletion of TiO could point towards condensation or cold trapping of TiO at the terminator as predicted by \cite{Parmentier2013}.

7) After accounting for different bin sizes, we obtain on average $\sim$40 \% better precision in the GMOS spectrum compared to the previous MOS optical transmission spectrum observed by \cite{Espinoza2019} and \cite{Sedaghati2017}. In terms of the spectral morphology, our non-detection of TiO features is consistent with the findings of \cite{Espinoza2019} but inconsistent with \cite{Sedaghati2017}. We tentatively detect Na absorption albeit at lower amplitude than that detected by \cite{Sedaghati2017}, which is inconsistent with \cite{Espinoza2019}. However, given the weak evidence for Na absorption in the Gemini/GMOS spectrum, we cannot definitively resolve the tension with respect to Na detection between \cite{Sedaghati2017} and \cite{Espinoza2019}. The Gemini/GMOS transmission spectrum is overall consistent with the sub-solar TiO scenario which is also reported by the previous studies from \cite{Espinoza2019} and \citeauthor{Sedaghati2017}, (\citeyear{Sedaghati2017}, \citeyear{Sedaghati2021}) using Magellan/IMACS, VLT/FORS2, and ESPRESSO respectively.

Our work ultimately demonstrates that multi-epoch transmission spectra from exoplanet transiting variable stars cannot be simply co-added or combined to improve the precision on the final transmission spectrum before correcting them for stellar variability effects. The method to correct for the effect of multi-epoch stellar variability introduced in this paper becomes even more relevant for high precision observations from JWST (\citealt{Zellem2017}, \citealt{Mayorga2021}). The issue for the near-infrared JWST observations will be most significant for active stars with high level of stellar variability. However, JWST NIRISS observations, which will go up to $\sim$600 nm in the blue optical, will have to correct for stellar variability effect before combining observations taken at different epochs.     

\section*{Acknowledgements}

Based on observations obtained at the Gemini Observatory (acquired through the Gemini Observatory Archive and Gemini Science Archive), which is operated by the Association of Universities for Research in Astronomy, Inc. (AURA), under a cooperative agreement with the NSF on behalf of the Gemini partnership: the National Science Foundation (United States), the National Research Council (Canada), CONICYT (Chile), Ministerio de Ciencia, Tecnolog\'{i}a e Innovaci\'{o}n Productiva (Argentina), Minist\'{e}rio da Ci\^{e}ncia, Tecnologia e Inova\c{c}\~{a}o (Brazil), and Korea Astronomy and Space Science Institute (Republic of Korea). Based in part on Gemini observations obtained from the National Optical Astronomy Observatory (NOAO) Prop. ID: 2012B-0398; PI: J-.M D\'{e}sert. We are very grateful to the anonymous reviewer for their careful and thorough feedback which greatly improved this work. V.P. is grateful to Filipe Matos for helping with the raw data reduction of the LCOGT light curves. V.P. acknowledges stimulating discussions with Lorenzo Pino and Jacob Arcangeli on ultra-hot Jupiters. V.P. acknowledges help from Ben Montet on using the \texttt{eleanor} package. J.M.D acknowledges support from the Amsterdam Academic Alliance (AAA) Program, and the European Research Council (ERC) European Union’s Horizon 2020 research and innovation programme (grant agreement no. 679633; Exo-Atmos). This work is part of the research programme VIDI New Frontiers in Exoplanetary Climatology with project number 614.001.601, which is (partly) financed by the Dutch Research Council (NWO). This material is based upon work supported by the NWO TOP Grant Module 2 (Project Number 614.001.601).  This material is based upon work supported by the National Science Foundation (NSF) under Grant No. AST-1413663. This research has made use of the NASA Exoplanet Archive, which is operated by the California Institute of Technology, under contract with the National Aeronautics and Space Administration under the Exoplanet Exploration Program. This research has made use of NASA's Astrophysics Data System. The authors also acknowledge the significant cultural role and reverence the summit of Mauna Kea has within the indigenous Hawaiian community. This research has made use of \texttt{Astropy},\footnote{http://www.astropy.org} a community-developed core Python package for Astronomy \cite{astropy:2013, astropy:2018}, \texttt{NumPy} \cite{harris2020array}, \texttt{matplotlib}  \cite{Hunter2007}, \texttt{SciPy} \cite{Virtanen2020} and \texttt{IRAF} \cite{Tody1986} distributed by the NOAO, which is operated by AURA under a cooperative agreement with the NSF.



\section*{Data Availability}
The data underlying this article and Python notebooks from which the results and figures of this paper can be obtained will be made available upon publication.







\appendix

\section{Photometric monitoring of WASP-19}
\label{app:photometric_monitoring}

\subsection{TESS observations of WASP-19b}
\label{tess_obs}

\begin{figure*}

  \centering
  \includegraphics[scale=0.5]{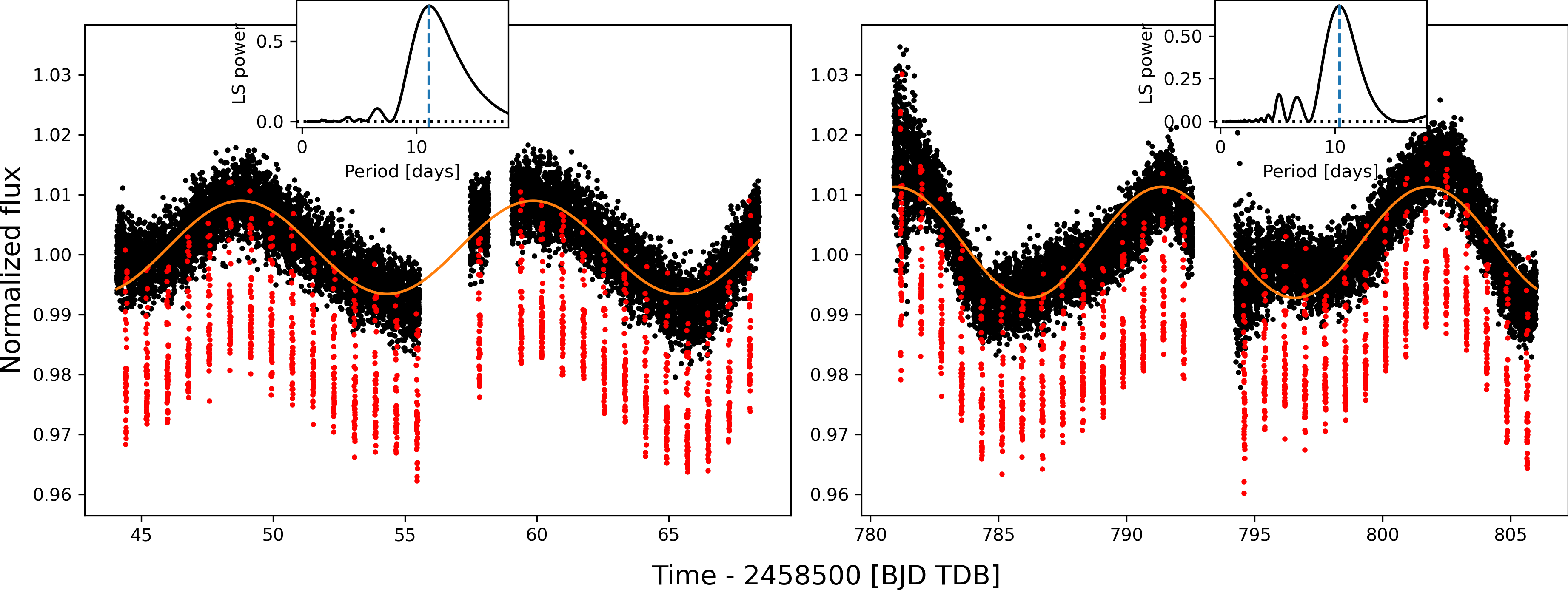}
  \caption[1]{TESS Simple Aperture Photometry (SAP) light curve of WASP-19 observed in Sector 9 (in the left panel) and Sector 36 (in the right panel) obtained from the latest version of the TESS SPOC pipeline from MAST. Both sectors have been corrected for dilution and bad quality exposures have been masked out (see Section \ref{tess_lc_analysis} for more detail). We have also normalized the two orbits for each sector by their respective median fluxes. The transits are marked by red points. The stellar flux varies by $\sim$ 2 \% peak to trough in both the sectors, with the Lomb Scargle periodogram (inset) of the out of transit flux times series showing a peak at 11.4 days for Sector 9 and 10.13 days for Sector 36. The orange line shows the sinusoidal fit to the photometry corresponding to the peak of the Lomb Scargle periodogram for each sector.   
  }
  \label{fig:tess_obs}
\end{figure*}

The TESS spacecraft observed WASP-19 (TIC 35516889, TOI 655) in Sector 9 (from 28 February 2019 to 26 March 2019) and Sector 36 (from 7 March 2021 to 2 April 2021) covering a total of 59 transits of WASP-19b (Figure \ref{fig:tess_obs}). TESS photometric observations are obtained in the broadband wavelength range of 600 to 1000 nm which overlaps with the GMOS-R150 wavelength range of 500 to 900 nm covered by our WASP-19 observations. Hence, TESS photometric light curves obtained over multiple epochs can be used to probe the effect of stellar variability on the broadband transit depth of WASP-19b in the GMOS bandpass. Moreover, the large number of transits observed by TESS can be used to benchmark the transit parameters of the system.    

For the analysis in this paper we obtained the Simple Aperture Photometry (SAP) TESS light curves for both sector 9 and 36 extracted using an optimal aperture size computed by the Science Processing Operations Center (SPOC) pipeline (\citealt{Jenkins2016}) and publicly available on the Mikulski Archive for Space Telescopes (MAST). As also previously noted by \cite{Wong2019} who used the Sector 9 data, the Presearch Data Conditioned (PDC) light curves made available by SPOC for WASP-19 introduce correlated features in the light curve which are not originally seen in the SAP light curve. 

A recent update to the TESS SPOC pipeline as described in \cite{Jenkins2021} implemented better sky subtraction which overcomes the overestimation of sky background which could lead to overestimation of the measured transit depth. MAST provides reduced light curves for Sector 9 from both the old and the latest version of the TESS SPOC pipeline, and for Sector 36 only from the latest version of the pipeline. We find that measured flux in the Sector 9 SAP light curves from the old and new version of the pipelines differ by on average 8 \% as shown in Figure 11 in the supplementary material. We also find that the light curves from the old TESS SPOC pipeline overestimate the transit depth of WASP-19b by 1500 ppm on average. Hence, we choose to use the SAP light curves from the latest version of the TESS SPOC pipeline for our analysis in this paper, which we describe in further detail. 

\subsubsection{Analysis of TESS Light Curves of WASP-19b}
\label{tess_lc_analysis}

The range of TESS photometry of WASP-19 is 27 days for individual sectors which is just over two stellar rotational cycles (P$_{rot}$ $\sim$ 10.5 days) and is evident as spot modulated stellar flux variability of $\sim$ 2 \% peak to trough. We first mask out the bad quality exposures using the one-hot encoded quality mask in the 'QUALITY' keyword in the header of the light curve files provided by SPOC (\citealt{Jenkins2016}). After masking out the bad quality exposures, one transit in sector 36 is masked out and hence we have 58 transits in total from both sectors for our final analysis. We then use the `CROWDSAP' keyword from the header for each sector light curve file to get an estimate of the ratio of target flux to total flux in optimal aperture used for the SAP photometry. This value can be used to subtract the dilution from nearby sources. The dilution flux we subtracted this way was 10.73 \% and 7.8 \% of the median measured flux for Sector 9 and Sector 36 respectively. 

We additionally clip any remaining outliers in SAP light curve at more than 3 $\sigma$ using a moving box average before fitting individual transit light curves.  We sliced the SAP light curves for WASP-19 into 58 individual transit light curves manually with approximately 4 hours before and after transit, to provide enough out of transit baseline to fit the transit signal. We then fit each light curve with the combined transit and a GP noise model similar to that used for the GMOS transit light curve in Section \ref{sec:wtlc} with time stamps of each exposure as a GP regressor. We fit for the orbital inclination ($i$), normalized orbital separation ($a/R_\star$), central transit time ($T_0$), planet to star radius ratio ($R_\mathrm{P}/R_\star$), and a linear limb darkening parameter. We used wide uniform priors for $R_\mathrm{P}/R_\star$, mid-transit time ($T_0$), semi major axis ($a/R_\star$), and inclination ($i$), and a Gaussian prior on the linear limb darkening coefficient with mean and variance fixed by the theoretically computed values from \texttt{PyLDTk} for the TESS bandpass (600 - 1000 nm). 

The best fit TESS light curves are shown in Figures 13 and 14 in the supplementary material. The weighted average value of inclination measured from both the sectors of TESS light curves is about 5$\sigma$ different from the most precise literature value reported by \cite{Espinoza2019}. In order to ensure that variation of the measured $a/R_\star$, and $i$ with each transit do not affect the measured transit depths, we also conducted light curve fits by fixing the inclination to the value measured by \cite{Espinoza2019}. We find that the transit depths from both the cases, whether we fit for the inclination or fix it, are consistent with each other within 1 $\sigma$ as seen in Figure 12 in the supplementary material.        

\subsubsection{TESS Light Curves of the comparison stars}
\label{tess_lc_analysis_comp_stars}
We also inspect the TESS light curves of the comparison stars observed by Gemini/GMOS in this paper. We obtained the TESS SPOC light curves for comparison star 2 (TIC 35516889) from MAST. Since the comparison star 1 (TIC 35516848) is faint, its light curves from the QLP pipeline (\cite{Huang2020}) available on MAST are significantly contaminated by the flux of WASP-19. Hence, we derived the light curves for the comparison star 1 by obtaining the full-frame images obtained from MAST and analysing them using the \texttt{eleanor} package (\citealt{Feinstein2019}). We show the light curves and the Lomb Scargle periodograms for both comparison stars in Figure 9 in the supplementary material. After independently normalizing the two TESS orbits within each sector to the median flux in the orbit, we find that the comparison star 1 shows  $\pm$ 0.2 \%  and $\pm$ 0.6 \% variability in Sector 9 and 36 respectively, while the comparison star 2 shows $\pm$ 0.1 \% and $\pm$ 0.2 \% variability in sector 9 and 36 respectively. From their Lomb Scargle periodograms, the light curves for comparison star 1 show periodicity at 11 and 10.74 days in sector 9 and 36 respectively as seen in Figure 11 in supplementary material. Similarly, comparison star 2 shows periodicity at 12.72 and 13.48 days for sector 9 and 36 respectively. For comparison star 2 this could represent the actual period of its variability as it is brighter than WASP-19. However, for the comparison star 1, which is fainter than both the other stars, we cannot rule out the possibility that the observed periodicity could stem from low level of contamination from either or both the other stars.  


  

  


\subsection{Ground based photometric monitoring of WASP-19 by LCOGT}
\label{lcogt_obs}
We obtained broadband long-term monitoring photometric observations of WASP-19 in the Johnson Cousins/Bessell B and R band using the Las Cumbres Observatory Global Telescope (LCOGT) network of robotic telescopes (\citealt{Brown2013}) to monitor the photometric variability of WASP-19 due to stellar activity. The LCO network consists of 42 telescopes with mirrors with diameters of 40 cm, 1 m, and 2 m spread across Earth in latitude and longitude, providing full sky coverage. We used the 40 cm and 1 m telescopes with the SBIG 4k x 4k to obtain 303 B-band photometric observations from 2014 May 5 to 2014 July 13, and 234 B-band and 190 R-band photometric observations from 2016 March 10 to 2017 January 26.

We used the scientific data outputs from the BANZAI and ORAC pipelines of LCOGT on which we further perform WCS correction, centroid fitting, and simple aperture photometry using a custom pipeline that uses modules from the Astropy package (\citealt{astropy:2013}, \citealt{astropy:2018}). For performing differential photometry we first select the comparison stars to use according to their magnitude in the B band, choosing the ones with $\Delta$B$_{mag}$ $\leq$ 1 and within 5” from WASP-19. We then choose the combination of comparison stars that results in a minimum scatter in the light curve with respect to the median. We do not notice any significant correlation between the differential flux and airmass for both seasons and choose not to perform any airmass correction. 

\begin{figure*}
  \centering
  \includegraphics[scale=0.6]{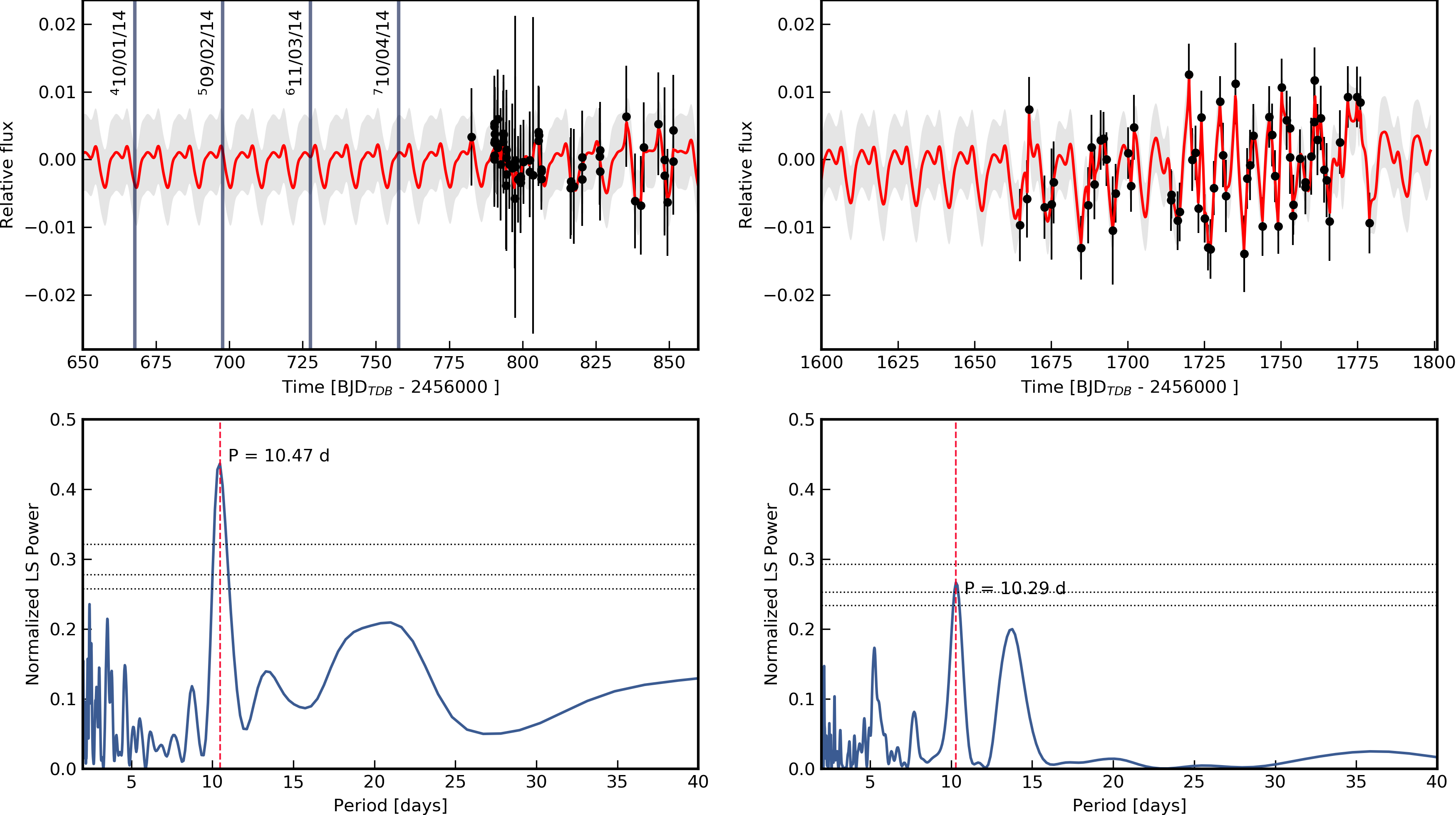}
  \caption[1]{ {\it Upper panel} : Photometric monitoring of WASP-19b obtained from the LCOGT network of telescopes in B band in two seasons (Upper left is Season 1, and upper right is Season 2). Both the seasons have been normalized with respect to their respective seasonal median. The red curve shows the best-fit obtained from the quasi-periodic Gaussian process regression model for both seasons combined which indicates quasi-periodic variations in stellar flux due to stellar rotation. Vertical lines mark the dates of our Gemini/GMOS observations (subscript indicating the observation number as specified in Table \ref{obsstats}). The period estimated from the best fit quasi-periodic Gaussian-process regression model is $P_{rot} = 10.91+^{0.1}_{0.09} days$ which is close to that found from the Lomb Scargle periodograms of the corresponding seasons and to that from TESS photometry in this paper and previous ground based photometry (\citealt{Espinoza2019}). The horizontal dotted lines in the LS periodograms in the {\it lower panel} indicate the 0.1, 0.05, and 0.01 False Alarm Probability levels, and the vertical dashed red line mark the location of maximum normalized power. 
  }
  \label{fig:LCO_phot}
\end{figure*}

The extracted LCOGT photometry of WASP-19 shows peak to trough variation of $\sim$ 2 \% during both seasons. We find period of flux variations as 10.47 and 10.29 days for the two seasons from their respective Lomb-Scargle periodograms as shown in Figure \ref{fig:LCO_phot}. We also fit the LCOGT photometry using a GP model with a quasi-periodic exponential sine-squared kernel implemented in \texttt{george} (\citealt{Ambikasaran2015}), which is also plotted in Figure \ref{fig:LCO_phot}. We find the `Period' hyperparameter from the GP fit to be $10.91+^{0.1}_{0.09} days$ which is consistent with the period of stellar flux variability measured from the Lomb-Scargle periodograms. 

\section{Illustration of the transmission spectral slope vs offset analysis}
\label{sec:slope_off_illustrate}
\begin{figure*}

  \centering
  \includegraphics[scale=0.3]{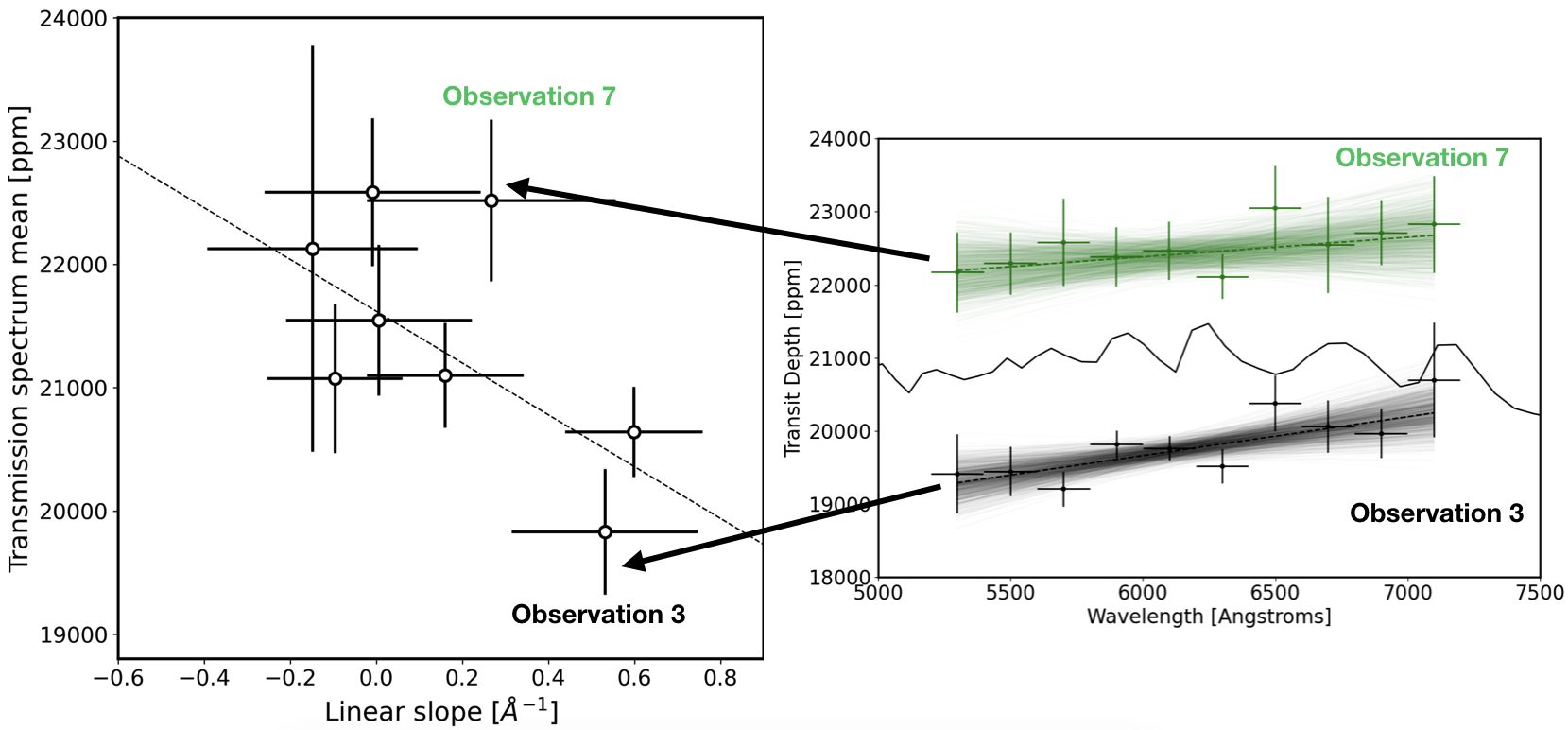}
  \caption[1]{Illustration of the construction of transmission spectra mean level vs slope space presented in Figure \ref{fig:TS_slope_vs_mean_level}. The black curve on the right panel shows \texttt{platon} transmission spectrum model without any stellar spot or faculae contribution and normalized to the HST/WFC3 spectrum from \cite{Huitson2013}. In black and green points on the right panel are shown the transmission spectra for observations 3 and 7 respectively. Overplotted are the linear fits to both the spectra along with randomly sampled fits from the \texttt{emcee} posterior for the fit. We plot the mean of the two spectra from observation 3 and 7 vs their measured linear slope on the left panel (shown by the black arrows). We repeat the process for all the eight epochs to populate the spectral mean vs slope space as shown here as black points in the left panel which is the same as Figure \ref{fig:TS_slope_vs_mean_level}. 
  }
  \label{fig:slope_off_illustrate}
\end{figure*}

In Figure \ref{fig:slope_off_illustrate} we visually illustrate how we construct the transmission spectra mean level vs slope space (shown in Figure \ref{fig:TS_slope_vs_mean_level}) to measure the relative effects of stellar variability on the transmission spectra obtained at different epochs.  

\section{Transmission spectra tables}

In this appendix we present the tables of transmission spectra obtained in Section \ref{sec:binned_LC}. Table \ref{tab:r150_ts_corr} shows the transmission spectra from the 8 transits obtained using the conventional method described in Section \ref{sec:ts_old} and Table \ref{tab:r150_ts_targ} shows the transmission spectra obtained using the new method in Section \ref{sec:ts_new}. The spectroscopic light curve fits from both the new and the conventional methods are shown in Figure 3 to 10 in the supplementary material. Table \ref{tab:comb_ts_targ} shows the median combined transmission spectrum after applying the slope and offset correction in Section \ref{sec:correct_stellar_activity}.     

\begin{table*}
\begin{center}
\caption[1]{Wavelength dependent transit depths (in ppm) for the individual GMOS-R150 observations (marked in the columns) obtained using the conventional method described in more detail in Section \ref{sec:ts_old}. }  
\label{tab:r150_ts_corr}
\begin{tabular}{ccccccccc}
\hline
Wavelength [\AA] & & & & Transit Depth [ppm] & & & \\ 
 & 1 & 2 & 3 & 4 & 5 & 6 & 7 & 8\\ 
\hline

5200 - 5400 & 19582 $\pm$ 505 & 20831 $\pm$ 766 & 19588 $\pm$ 570 & 18685 $\pm$ 497 & 19576 $\pm$ 552 & 21251 $\pm$ 497 & 21143 $\pm$ 767 & - \\
5400 - 5600 & 18694 $\pm$ 364 & 20821 $\pm$ 275 & 19443 $\pm$ 276 & 19269 $\pm$ 287 & 19924 $\pm$ 371 & 20522 $\pm$ 901 & 22418 $\pm$ 447 & 18876 $\pm$ 986 \\
5600 - 5800 & 19781 $\pm$ 250 & 20913 $\pm$ 255 & 19232 $\pm$ 220 & 19022 $\pm$ 267 & 20197 $\pm$ 237 & 21330 $\pm$ 287 & 21091 $\pm$ 363 & 19064 $\pm$ 743 \\
5800 - 6000 & 19975 $\pm$ 257 & 20712 $\pm$ 228 & 19386 $\pm$ 237 & 19620 $\pm$ 216 & 19839 $\pm$ 197 & 20587 $\pm$ 342 & 20123 $\pm$ 1049 & 17760 $\pm$ 765 \\
6000 - 6200 & 19890 $\pm$ 278 & 21453 $\pm$ 349 & 19890 $\pm$ 191 & 19620 $\pm$ 199 & 20082 $\pm$ 242 & 20919 $\pm$ 324 & 21557 $\pm$ 377 & 18060 $\pm$ 975 \\
6200 - 6400 & 20047 $\pm$ 240 & 21329 $\pm$ 213 & 19773 $\pm$ 169 & 19326 $\pm$ 203 & 20260 $\pm$ 270 & 20614 $\pm$ 511 & 20786 $\pm$ 508 & 16215 $\pm$ 935 \\
6400 - 6600 & 19672 $\pm$ 452 & 20805 $\pm$ 211 & 19063 $\pm$ 1042 & 19465 $\pm$ 298 & 20344 $\pm$ 173 & 20785 $\pm$ 248 & 20972 $\pm$ 465 & 16657 $\pm$ 953 \\
6600 - 6800 & 19611 $\pm$ 212 & 21082 $\pm$ 222 & 19470 $\pm$ 190 & 19652 $\pm$ 255 & 20241 $\pm$ 174 & 20709 $\pm$ 521 & 20631 $\pm$ 543 & 19570 $\pm$ 1445 \\
6800 - 7000 & 19708 $\pm$ 271 & 21491 $\pm$ 309 & 19945 $\pm$ 196 & 19336 $\pm$ 799 & 19794 $\pm$ 207 & 20500 $\pm$ 611 & 21142 $\pm$ 428 & 18182 $\pm$ 683 \\
7000 - 7200 & 19696 $\pm$ 212 & 20714 $\pm$ 1033 & 19391 $\pm$ 169 & 19161 $\pm$ 601 & 20241 $\pm$ 165 & 20312 $\pm$ 196 & 21215 $\pm$ 464 & 19424 $\pm$ 647 \\
7200 - 7400 & - & - & - & - & - & - & - & 19664 $\pm$ 543 \\
8199 - 8401 & - & - & - & - & - & - & - & 19921 $\pm$ 692 \\
8401 - 8600 & - & - & - & - & - & - & - & 19860 $\pm$ 836 \\
8600 - 8800 & - & - & - & - & - & - & - & 18842 $\pm$ 711 \\
8800 - 8999 & - & - & - & - & - & - & - & 19345 $\pm$ 900 \\

\hline

\end{tabular}
\end{center}
\end{table*}

\begin{table*}
\begin{center}

\caption[1]{Wavelength dependent transit depths (in ppm) for the individual GMOS-R150 observations (marked in the columns) obtained using the new method introduced by \citetalias{Panwar2022} and described in Section \ref{sec:ts_new}. }  
\label{tab:r150_ts_targ}
\begin{tabular}{ccccccccc}
\hline
Wavelength [\AA] & & & & Transit Depth [ppm] & & & \\ 
 & 1 & 2 & 3 & 4 & 5 & 6 & 7 & 8\\ 
\hline

5200 - 5400 & 20414 $\pm$ 983 & 21446 $\pm$ 542 & 19418 $\pm$ 541 & 20228 $\pm$ 483 & 23945 $\pm$ 922 & 22099 $\pm$ 1079 & 22173 $\pm$ 547 & 22913 $\pm$ 1486 \\
5400 - 5600 & 20927 $\pm$ 250 & 20901 $\pm$ 364 & 19451 $\pm$ 337 & 20057 $\pm$ 226 & 22365 $\pm$ 482 & 21217 $\pm$ 378 & 22294 $\pm$ 423 & 23280 $\pm$ 1287 \\
5600 - 5800 & 21026 $\pm$ 335 & 20842 $\pm$ 297 & 19211 $\pm$ 240 & 20365 $\pm$ 316 & 22291 $\pm$ 391 & 21335 $\pm$ 371 & 22583 $\pm$ 596 & 23100 $\pm$ 957 \\
5800 - 6000 & 21431 $\pm$ 87 & 20798 $\pm$ 205 & 19819 $\pm$ 187 & 20468 $\pm$ 249 & 22309 $\pm$ 440 & 21735 $\pm$ 313 & 22387 $\pm$ 409 & 22409 $\pm$ 714 \\
6000 - 6200 & 20916 $\pm$ 401 & 21402 $\pm$ 205 & 19769 $\pm$ 167 & 20363 $\pm$ 263 & 22488 $\pm$ 179 & 21984 $\pm$ 414 & 22468 $\pm$ 395 & 21279 $\pm$ 769 \\
6200 - 6400 & 20999 $\pm$ 218 & 21135 $\pm$ 200 & 19523 $\pm$ 235 & 21150 $\pm$ 176 & 22192 $\pm$ 309 & 21018 $\pm$ 301 & 22113 $\pm$ 305 & 21682 $\pm$ 460 \\
6400 - 6600 & 21640 $\pm$ 815 & 20833 $\pm$ 286 & 20381 $\pm$ 382 & 21158 $\pm$ 208 & 22189 $\pm$ 177 & 21402 $\pm$ 161 & 23054 $\pm$ 578 & 21691 $\pm$ 283 \\
6600 - 6800 & 20929 $\pm$ 297 & 21341 $\pm$ 236 & 20066 $\pm$ 357 & 20478 $\pm$ 268 & 22362 $\pm$ 357 & 21926 $\pm$ 389 & 22547 $\pm$ 656 & 21727 $\pm$ 258 \\
6800 - 7000 & 21125 $\pm$ 227 & 21146 $\pm$ 206 & 19969 $\pm$ 330 & 20842 $\pm$ 222 & 23154 $\pm$ 410 & 21474 $\pm$ 229 & 22714 $\pm$ 438 & 21804 $\pm$ 132 \\
7000 - 7200 & 21336 $\pm$ 294 & 21143 $\pm$ 516 & 20700 $\pm$ 783 & 21289 $\pm$ 301 & 22558 $\pm$ 558 & 21279 $\pm$ 499 & 22832 $\pm$ 661 & 22478 $\pm$ 699 \\
7200 - 7400 & - & - & - & - & - & - & - & 22605 $\pm$ 1528 \\
8199 - 8401 & - & - & - & - & - & - & - & 21162 $\pm$ 1799 \\
8401 - 8600 & - & - & - & - & - & - & - & 21464 $\pm$ 1078 \\
8600 - 8800 & - & - & - & - & - & - & - & 22006 $\pm$ 1199 \\
8800 - 8999 & - & - & - & - & - & - & - & 22312 $\pm$ 1449 \\

\hline

\end{tabular}
\end{center}
\end{table*}

\begin{table}
\begin{center}

\caption[1]{Combined transmission spectrum obtained after applying the wavelength-dependent offset (slope) correction to the spectrum at each epoch and then weighted median combining them, as described in more detail in Section \ref{sec:correct_stellar_activity}.}  
\label{tab:comb_ts_targ}
\begin{tabular}{cc}
\hline
Wavelength [\AA] & Transit Depth [ppm] \\ 
\hline
5200 - 5400 & 21812 $\pm$ 236 \\
5400 - 5600 & 21477 $\pm$ 120 \\
5600 - 5800 & 21522 $\pm$ 124 \\
5800 - 6000 & 21831 $\pm$ 67 \\
6000 - 6200 & 21745 $\pm$ 89 \\
6200 - 6400 & 21497 $\pm$ 87 \\
6400 - 6600 & 21626 $\pm$ 88 \\
6600 - 6800 & 21664 $\pm$ 110 \\ 
6800 - 7000 & 21690 $\pm$ 79 \\
7000 - 7200 & 21842 $\pm$ 159 \\

\hline

\end{tabular}
\end{center}
\end{table}


\bsp	
\label{lastpage}
\end{document}